\documentclass[sigconf, nonacm]{acmart}

\usepackage{makecell}
\usepackage{microtype}

\AtBeginDocument{%
  }

\usepackage{color}
\usepackage{soul}    
\usepackage{gensymb} 
\usepackage{subcaption}
\usepackage[dvipsnames]{xcolor}
\usepackage{graphicx}
\usepackage[inline]{enumitem} 
\usepackage{tabularray}
\usepackage{listings}
\usepackage{gensymb} 
\usepackage{xspace} 
\usepackage{pifont}
\usepackage{multirow}
\usepackage{setspace}

\newcommand{\cmark}{\ding{51}}%
\newcommand{\xmark}{\ding{55}}%

\newcommand{\systemName}{\textsc{SimAB}\xspace} 

\lstdefinestyle{prompt}{
  basicstyle=\ttfamily\small\color{black}\setstretch{0.8},
  showstringspaces=false,
  breaklines=true
}
\definecolor{mybrown}{HTML}{7B241C}
\definecolor{myblue}{HTML}
{0E6655}
\definecolor{mygray}{HTML}{595959}

\usepackage{cleveref}

\begin{document}

\title[SimAB: Simulating A/B Tests with Persona-Conditioned AI Agents for Rapid Design Evaluation]{\systemName: Simulating A/B Tests with Persona-Conditioned AI Agents for Rapid Design Evaluation}

\author{Tim Rieder}
\authornote{Equal contribution. The author ordering was randomized.}
\authornote{Work done during an internship at Adobe Research.}
\affiliation{\institution{ETH Zurich}\city{Zurich}\country{Switzerland}}
\email{timrieder@ethz.ch}

\author{Marian Schneider}
\authornotemark[1]
\authornotemark[2]
\affiliation{\institution{ETH Zurich}\city{Zurich}\country{Switzerland}}
\email{smarian@ethz.ch}


\author{Mario Truss}
\affiliation{\institution{Adobe}\city{Munich}\country{Germany}}
\email{mtruss@adobe.com}

\author{Vitaly Tsaplin}
\affiliation{\institution{Adobe}\city{Basel}\country{Switzerland}}
\email{tsaplin@adobe.com}

\author{Alina Rublea}
\affiliation{\institution{Adobe}\city{Basel}\country{Switzerland}}
\email{rublea@adobe.com}

\author{Sinem Dere}
\affiliation{\institution{Adobe}\city{Basel}\country{Switzerland}}
\email{ndere@adobe.com}

\author{Francisco Chicharro Sanz}
\affiliation{\institution{Adobe}\city{Basel}\country{Switzerland}}
\email{chicharr@adobe.com}

\author{Tobias Reiss}
\affiliation{\institution{Adobe}\city{Basel}\country{Switzerland}}
\email{reiss@adobe.com}

\author{Mustafa Doga Dogan}
\affiliation{\institution{Adobe Research}\city{Basel}\country{Switzerland}}
\email{doga@adobe.com}

\renewcommand{\shortauthors}{Rieder and Schneider et al.}

\begin{abstract}

A/B testing is a standard method for validating design decisions, yet its reliance on real user traffic limits iteration speed and makes certain experiments impractical. We present \systemName, a system that reframes A/B testing as a fast, privacy-preserving simulation using persona-conditioned AI agents. Given webpage design screenshots and a conversion goal, \systemName generates user personas, deploys them as agents that state their preference, aggregates results, and synthesizes rationales.
Through a formative study with experimentation practitioners, we identified scenarios where traffic constraints hinder testing, including low-traffic pages, multi-variant comparisons, micro-optimizations, and privacy-sensitive contexts. Our design emphasizes speed, early feedback, actionable rationales, and audience specification. We evaluate \systemName against 47 historical A/B tests with known outcomes testing webpage layout and content, achieving 67\% overall accuracy, increasing to 83\% for high-confidence cases. Additional experiments show robustness to naming and positional bias and demonstrate accuracy gains from personas. Practitioner feedback suggests that \systemName supports faster evaluation cycles and rapid screening of designs difficult to assess with traditional A/B tests.

\end{abstract}


\begin{teaserfigure}
  \centering
  \includegraphics[width=\textwidth]{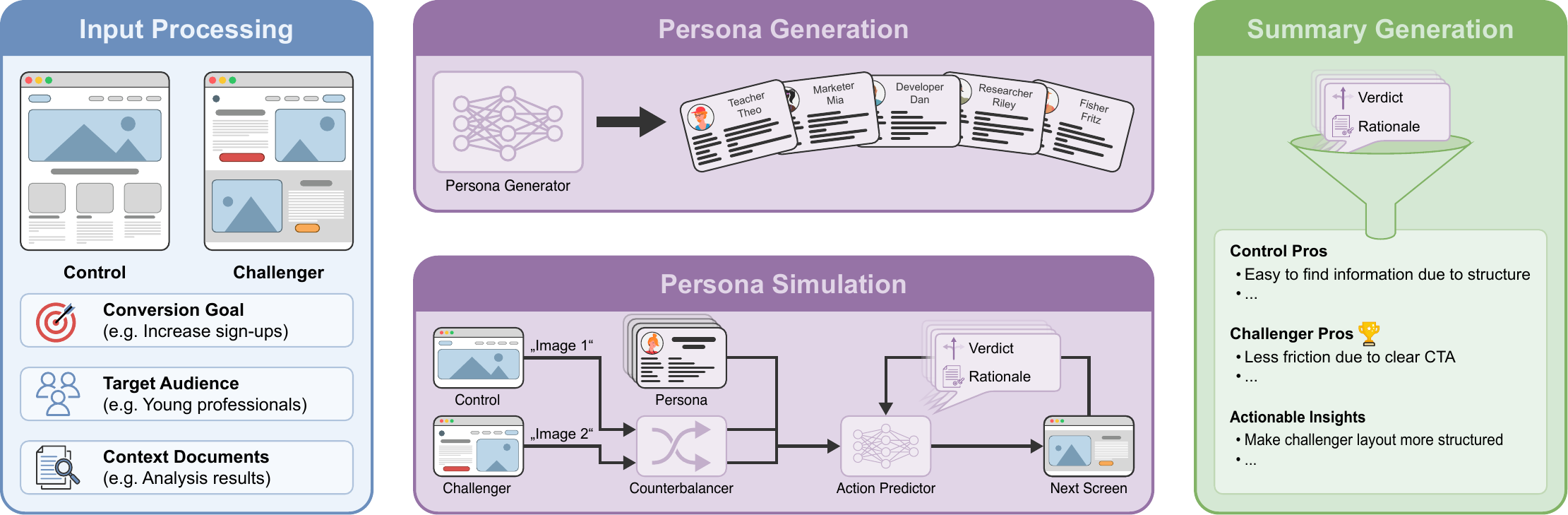}
  \caption{The \systemName pipeline reframes A/B testing as a fast simulation. 
  (1) \textbf{Input Processing}: The system ingests a test specification, containing design variants, conversion goals, target audience definitions, and documents to provide context. 
  (2) \textbf{Persona Generation}: It creates a diverse population of synthetic users (e.g., ``Marketer Mia'') conditioned on the specific context. 
  (3) \textbf{Persona Simulation}: Persona-driven agents interact with the designs. 
  (4) \textbf{Summary Generation}: Individual verdicts and rationales are aggregated into actionable insights. 
  This workflow reduces feedback latency from months to minutes.
  }
  \Description{A pipeline diagram showing four stages. 1. Input Processing shows Control and Challenger screenshots with goals. 2. Persona Generation shows a neural network generating persona cards like 'Teacher Theo' and 'Marketer Mia'. 3. Persona Simulation shows the agents interacting with the designs via a counterbalancer and action predictor. 4. Summary Generation shows a funnel aggregating verdicts into pros/cons and actionable insights.}
  \Description{The figure illustrates the SimAB pipeline for simulated A/B testing, organized into four stages from left to right.
Input Processing: Two design variants, labeled Control and Challenger, are shown as webpage screenshots. Additional inputs include a conversion goal, such as increasing sign-ups, a target audience definition, such as young professionals, and optional context documents, such as analysis results.
Persona Generation: A persona generator model creates multiple synthetic user personas. Example persona cards include different roles and backgrounds, such as teacher, marketer, developer, researcher, and others, representing diverse user perspectives.
Persona Simulation: Each persona evaluates both design variants. The variants are labeled neutrally as “Image 1” and “Image 2” and passed through a counterbalancer to mitigate ordering bias. Personas interact with the designs through an action predictor that simulates navigation and next-screen behavior. Each persona produces a verdict and a rationale.
Summary Generation: Individual verdicts and rationales are aggregated into a summary. The output includes a final verdict, grouped pros for the control and challenger designs, and a list of actionable design insights.}
  \label{fig:teaser}
\end{teaserfigure}


\maketitle

\section{Introduction}
\label{sec:introduction}

A/B testing has established itself as a gold standard for estimating causal effects of design decisions. It offers a data-driven method to measure the impact of changes by simultaneously deploying the original (\textit{Control}) and a modified (\textit{Challenger}) variant to real users to obtain user behavior data for both variants. Based on the collected data, the better variant can be identified according to metrics, such as click-through rate and annual recurring revenue. This methodology is commonly used in web development, where traffic can be easily routed to different variants to quantify how design changes impact performance metrics~\citep{kohavi_online_2013}.

While very effective, the reliance on real user data creates a significant bottleneck. Reaching statistically significant results requires large amounts of user traffic, often requiring months to collect, which is especially severe for pages with low traffic, that might never reach statistical significance. Furthermore, traditional testing validates ideas only at the end of the development cycle. This implies that significant effort is expended on variants that effectively yield no return if the test fails. These limitations highlight the need for a fast evaluation environment that makes testing accessible without the reliance on real users.

Recent advances in large language models (LLMs) have sparked interest in using these models to simulate human behavior~\citep{park_generative_2023, argyle_out_2023, horton_large_2023}. These \emph{synthetic audience} approaches promise to generate predictions about    design variants without waiting for real users. However, existing approaches either lack rigorous validation efforts~\citep{agent_ab_2024} or are limited to critiquing designs, instead of validating them~\citep{shinPosterMateAudiencedrivenCollaborative2025}.

In this work, we address these limitations by presenting \systemName, a system that leverages scalable, persona-conditioned, and interactive AI agents to generate synthetic preference judgments, reducing the time necessary to reach statistical significance to minutes. \systemName generates diverse user personas based on provided design variants and a test specification, deploys these personas as AI agents that evaluate design variants, and aggregates their judgments. We acknowledge systematic biases in LLMs, such as naming and positional bias~\citep{zheng_judging_2023}, and employ counterbalancing and neutral naming to mitigate them. We conduct extensive validation efforts using a custom corpus of 47 historical A/B tests to assess \systemName's accuracy, evaluate the effectiveness of our persona-conditioning and bias mitigation methods, and conduct feedback sessions with practitioners to validate the effectiveness of \systemName in real-world settings.

\vspace{0.1cm}
In summary, we make the following contributions:
\vspace{-0.1cm}
\begin{enumerate}
    \item \textbf{A formative study} (\Cref{sec:formative}) identifying the need for a rapid A/B testing system through four pain points and deriving the design implications of the system.
    
    \item \textbf{A synthetic A/B testing system} (\Cref{sec:system}) leveraging persona-conditioned agents to allow rapid A/B testing.
    
    \item \textbf{Comprehensive empirical validation} (\Cref{sec:evaluation}) demonstrating 67\% accuracy against 47 historical A/B tests, improving to $>$80\% when considering high-certainty tests.
\end{enumerate}

\section{Related Work}
\label{sec:related_work}

In this section, we discuss prior work that informs our approach to validating and optimizing designs without reliance on extensive real user traffic. We review research on A/B testing and optimization platforms, LLM-based design feedback and evaluation, and user-simulating agents, and we situate our contribution relative to their capabilities and limitations.

\subsection{A/B Testing and Optimization Platforms}

A/B testing has been the cornerstone of data-driven design optimization since its popularization in web contexts~\citep{kohavi_online_2013}. While algorithms such as multi-armed bandits~\citep{slivkins_introduction_2019} and Thompson sampling~\citep{gopalan_thompson_2014,zhao_thompson_ctr_2025} reduce the traffic required by dynamically allocating users to better-performing variants, these optimizations address \emph{efficiency} rather than the fundamental \emph{dependency on real user traffic}. Recent work has explored the simulation of user traffic via statistical frameworks~\citep{agarwal_synthetic_ab} and LLM-based approaches for market research~\citep{brand_gpt_market_2024}. However, these still require initial real-world observations or face limitations in generating externally valid consumer insights, limiting their applicability and effectiveness.

\subsection{LLMs for Design Feedback and Evaluation}

LLM-based approaches have been explored for design feedback and UI evaluation~\citep{ding_designgpt_2023,duanGeneratingUIDesign2023,duanGeneratingAutomaticFeedback2024,duanUICritEnhancingAutomated2024,duanVisualPromptingIterative2025,schmidt_simulating_hcd_2024,zheng_evalignux_2025,benharrakWriterDefinedAIPersonas2024,choiProxonaSupportingCreators2025,shinPosterMateAudiencedrivenCollaborative2025,suhStoryEnsembleEnablingDynamic2025}. Prior work has developed tools such as Figma plugins that use GPT-4 to automate heuristic evaluation of UI mockups~\citep{duanGeneratingAutomaticFeedback2024}, finding that while GPT-4 performs well on poor UI designs and is useful for catching errors and improving text, its performance decreases over iterations as designs improve, making it less suitable for iterative use by expert designers. 

Datasets have been created to improve LLM-based automated UI evaluation, with targeted datasets achieving 55\% performance gains through few-shot and visual prompting techniques~\citep{duanUICritEnhancingAutomated2024}. More recent work has proposed multimodal approaches using visual prompting and iterative refinement that reduce the gap from human performance by 50\% for UI critique generation~\citep{duanVisualPromptingIterative2025}, and these advancements have been applied to UX evaluation~\citep{zheng_evalignux_2025}. Persona-based approaches have been explored to generate on-demand feedback~\citep{benharrakWriterDefinedAIPersonas2024,choiProxonaSupportingCreators2025,shinPosterMateAudiencedrivenCollaborative2025} and support design iteration~\citep{suhStoryEnsembleEnablingDynamic2025}, and to derive design requirements from qualitative interview data~\citep{depaoliUserPersonasIdeation2026}. However, these approaches essentially model the behavior of a UX expert critiquing an interface, rather than a user trying to achieve a goal. While useful for spotting standard usability flaws, they fail to capture the complex, often irrational preferences real users can have.

\subsection{User Simulating Agents}

Based on these foundational usages of LLMs for design feedback, researchers have explored simulating actual users and their behaviors. Prior work has demonstrated that LLM agents can navigate real websites~\citep{he_webvoyager_2024} and established benchmarks for autonomous web agents~\citep{zhou_webarena_2024, lu_multiturn_2025}. 
Agent-based systems have been proposed for A/B testing~\citep{wang_agentab_2025} and usability testing~\citep{lu_uxagent_2025}, although their alignment with actual human behavioral responses to design interventions remains underexplored.

The use of AI to simulate human behavior has gained significant traction. Prior work has introduced \emph{generative agents} that simulate believable human behavior through memory retrieval and action planning~\citep{park_generative_2023}, with subsequent work demonstrating generative agent simulations of real individuals that replicate survey responses with high accuracy~\citep{park_generative_agents_1000}, finetuning approaches that improve simulation accuracy~\citep{kolluri_socrates_2025}, and large-scale simulations advancing understanding of collective behaviors~\citep{piao_agentsociety_2025}. Early work explored creating populated prototypes for social computing systems~\citep{parkSocialSimulacraCreating2022} and generating synthetic human-computer-interaction (HCI) research data~\citep{hamalainenEvaluatingLargeLanguage2023}. Researchers have explored LLMs as proxies for human participants, coining terms such as ``algorithmic fidelity''~\citep{argyle_out_2023} and ``silicon samples''~\citep{horton_large_2023,sarstedt_silicon_2024}, surveying user modeling approaches~\citep{tan_user_modeling_2023}, and introducing ``Synthetic User'' personas~\citep{gu_synthetic_users_2025}. Building on foundational work on personas~\citep{pruittPersonasPracticeTheory2003}, prior work has investigated how people customize and interact with agent personas in LLMs~\citep{haCloChatUnderstandingHow2024}. The digital twins paradigm has emerged as a powerful extension, with work introducing large-scale datasets~\citep{toubia_twin2k500_2025} and evaluating digital twins across multiple studies, finding that detailed personal information improves correlation beyond demographics-only personas~\citep{peng_megastudy_2025}.

Prior work has demonstrated LLM-based agents simulating complex multi-turn user behaviors for recommendation systems~\citep{wang_user_behavior_2025,zhang_agentcf_2023,zhang_generative_agents_rec_2023,chen_recusersim_2025}, with e-commerce applications including datasets of shopping behaviors~\citep{wang_opera_2025}, persona-aligned frameworks~\citep{mansour_paars_2025}, comparisons with human participants~\citep{sun_llm_agent_agentic_2025}, and reinforcement learning approaches for personalization~\citep{wang_customerr1_2025}. For persona development and evaluation, prior work has systematically reviewed GenAI approaches~\citep{amin_genai_personas_2025}, proposed methods for generating personas aligned with population distributions~\citep{hu_population_aligned_2025}, introduced benchmarks for persona consistency~\citep{samuel_personagym_2024}, developed trainable agents for role-play\-ing~\citep{shao_characterllm_2023}, created datasets for evaluation~\citep{buck_blueprint_2025}, studied human-AI workflows for collaborative persona generation~\citep{shinUnderstandingHumanAIWorkflows2024}, and addressed consistency through multi-turn reinforcement learning~\citep{abdulhai_consistent_personas_2025}.

However, significant concerns remain about the validity of synthetic data for understanding human preferences. Prior work has shown that LLM outputs reflect perspectives of specific demographic groups~\citep{santurkar_whose_2023}, identified systematic failures in capturing nuanced preferences~\citep{zhu_reliable_simulator_2024}, quantified behavioral gaps between humans and agents~\citep{wang_human_vs_agent_2025}, found limitations in GenAI-based usability testing~\citep{pourasad_genai_usability_2024}, and highlighted challenges in validating simulated behaviors~\citep{chen_towards_2025}. Systematic reviews have raised validity and reproducibility concerns about LLM use in HCI research~\citep{pangUnderstandingLLMificationCHI2025}, and studies have shown that LLMs replacing human participants can harmfully misportray and flatten identity groups~\citep{wangLargeLanguageModels2025}. Motivated by this, we conduct a comprehensive evaluation (\Cref{sec:evaluation}) of our proposed system to ensure that it provides reliable outputs.

In summary, existing approaches either depend on real user traffic (traditional A/B testing) or provide open-ended qualitative critiques instead of validation. \systemName distinguishes itself by leveraging synthetic audiences within a structured aggregation framework that enables direct pairwise comparison of designs. Unlike prior agent-based explorations, we move beyond proof-of-concept simulations by integrating systematic bias mitigation and providing the first extensive empirical validation against a large corpus of historical A/B tests with known ground-truth outcomes.

\section{Formative Study}
\label{sec:formative}

We grounded the design of \systemName in a formative study consisting of a 
qualitative synthesis of practitioner challenges derived from sustained collaboration with experimentation teams.

\subsection{Informative Interviews with Practitioners} \label{sub:formative_study_practitioners}

Based on participatory design engagements with 14 practitioners across four enterprise organizations, we identified key contextual factors that shaped the design of \systemName.
These organizations span diverse sectors, including telecommunications, industrial equipment manufacturing, and software distribution. All participating companies operated mature experimentation programs and were responsible for high-traffic software products and marketing pages. The A/B testing practices in the participating organizations were established for up to 15+ years.
The participants (10 men, 4 women) held senior roles ranging from product marketing specialists to principal product managers and senior managers of digital experience platforms. This cohort possessed significant domain expertise, reporting an average of 14 years of marketing experience (max 20+ years) and an average of 5.8 years specifically in A/B testing (max 10+ years).
We engaged in 1-5 sessions of 30 minutes of online meetings with each of those organizations and also received written feedback on personalized test results we provided.
We collected detailed notes from these sessions, which we iteratively synthesized to extract recurring challenges and design implications.

In the following, we describe both the pain-points our tool should address (\Cref{sub:customer_pain_points}) as well as key design principles that the tool should follow whilst addressing these pain-points (\Cref{sub:design_implications}).

\subsection{A/B Testing Pain Points} \label{sub:customer_pain_points}

We identified four primary scenarios where traditional A/B testing becomes costly or infeasible:

\textbf{Low-Traffic Pages}: Many pages, in emerging markets, serving niche audiences or business-to-business (B2B) segments, lack the traffic to reach statistical significance in a reasonable timeframe. According to industry estimates, 70-80\% of proposed tests are abandoned before reaching significance due to insufficient traffic \cite{Sharma2025ABTesting}.

\textbf{High Volume of Design Variants}: When the number of design variants is large, exhaustive A/B/n testing becomes infeasible because the required traffic grows rapidly with the number of variants (linearly, and often super-linearly)~\cite{Kaufmann2016BestArm,Jamieson2014BestArmSurvey}. Consequently, practitioners feel forced to limit their creativity, testing only one or two conservative changes rather than exploring diverse design hypotheses, simply to ensure the test concludes within a business cycle.

\textbf{Low-Value Changes}: Small changes are often not tested because the expected lift does not justify the cost and time of a full A/B test. Practitioners know of multiple tests to run, knowing they have an impact and lift, but they are too marginal to justify running them.

\textbf{Instrumentation \& Sampling Constraints}: Even on high-traffic pages, capturing valid human feedback is increasingly difficult. Practitioners noted that privacy regulations often limit data collection to users who opt in to cookies, creating a biased subset that may not represent the general population. 

These pain points create a barrier that frequently forces teams to abandon testing entirely. As a result, practitioners often deploy changes based on intuition rather than data, despite recognizing the necessity of validation.

\subsection{Design Implications \& Principles} \label{sub:design_implications}

To address these challenges and effectively integrate into practitioner workflows, we derived four key design principles to guide the development of \systemName.

\begin{enumerate}[leftmargin=*]
    \item \textbf{Shift-Left in the Design Process}: Traditional experimentation feedback arrives too late in the development cycle. Even if live results were instantaneous, they occur at the end of a long design pipeline, after substantial investment in ideation, implementation, and iteration. 
    To provide meaningful value, \systemName therefore needed to support predictive screening earlier in the workflow, before full implementation.

    \item \textbf{Prioritize Speed Over Perfect Accuracy}: Given that the alternative to synthetic A/B testing is often no testing, practitioners prefer fast, directional   guidance. The system should reach conclusions in minutes, not days, functioning as a rapid screening mechanism that filters distinct failures before they reach live traffic.
    
    \item \textbf{Audience Restrictions \& Additional Context}: Long-time A/B testing practitioners have prior knowledge of their (target) audience and want to leverage this to better specify the audience and their behavior (as potentially retrieved from finished A/B tests) accordingly. We additionally see an appealing use case, 
    where the user can restrict the audience to test how a product would land in a new market using the same mechanism, but this feature was secondary to leveraging prior knowledge from old A/B-tests to improve accuracy on the new tests.
    
    \item \textbf{Rationales \& Actionable Insights}:
    A major limitation of traditional A/B testing is that it reveals \emph{which} variant won, but determining the  \emph{why} is infeasible. To provide long-term value and allow for fast iteration, \systemName must be interpretable, moving beyond binary winners to ensure that the system functions as an explanatory tool rather than a black box.
\end{enumerate}

\section{System Design}
\label{sec:system}

Based on insights from the formative study, we designed \systemName, a system to test, validate, and iterate on early-stage designs using persona-conditioned AI agents simulating user interaction with and preference between design variations. The pipeline, visualized in \Cref{fig:teaser} consists of four stages: 
\begin{enumerate*}
    \item input processing and context retrieval,
    \item persona generation,
    \item persona simulation, and
    \item aggregation and summary generation.
\end{enumerate*}
We describe each stage, emphasizing design decisions informed by our formative study and validation experiments. We will conclude each of the following introductory subsections with a pictorial walkthrough of the complete flow, whereby an A/B test from the Wikimedia Foundation\footnote{\href{https://meta.wikimedia.org/wiki/Fundraising_2011/Test_Updates/October/21}{https://meta.wikimedia.org/wiki/Fundraising\_2011/Test\_Updates/October/21}} serves as an example.

\subsection{Input Processing}

\begin{figure*}[t]
  \centering
  \includegraphics[width=0.75\textwidth]{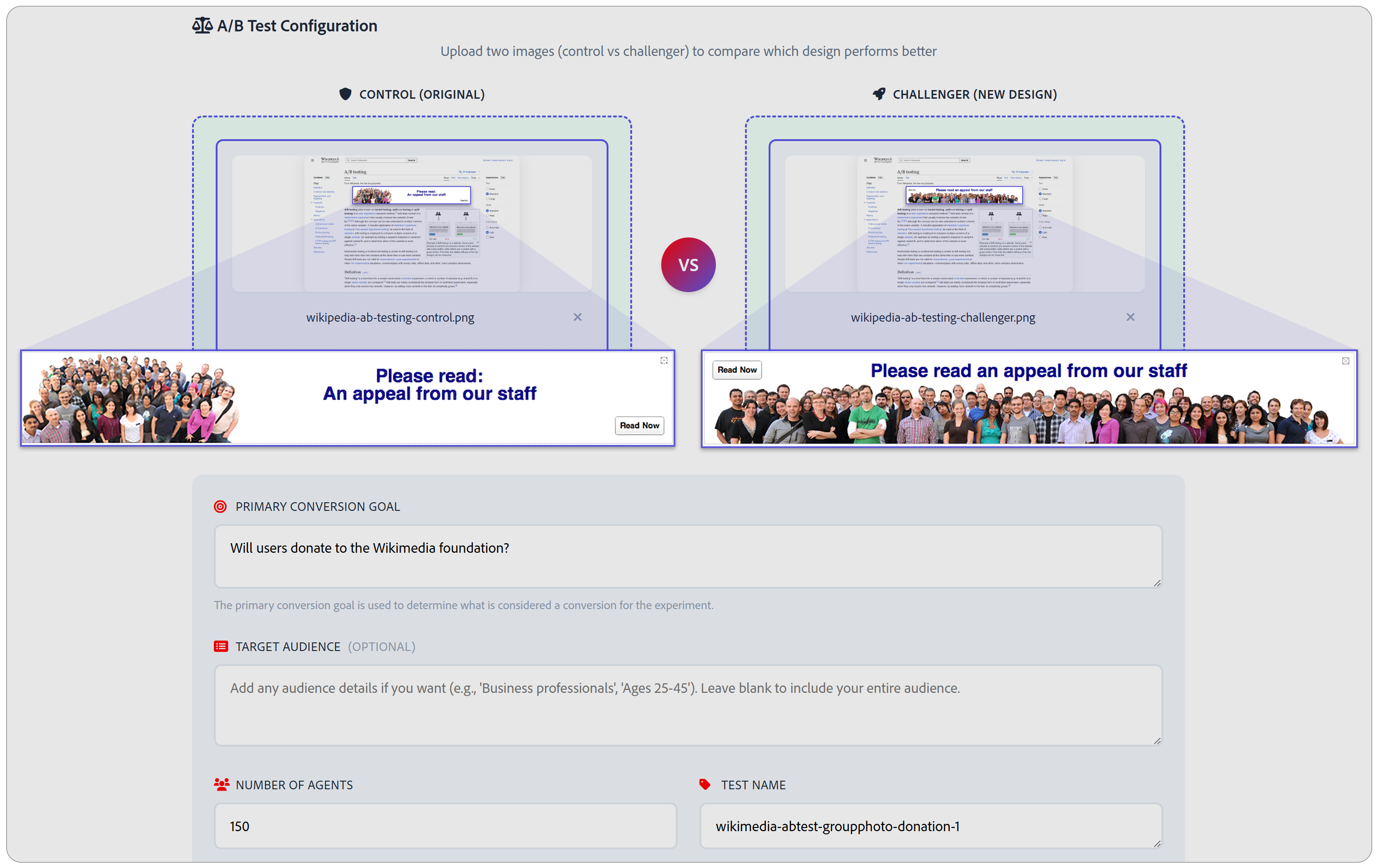}
  \caption{Example Input. Control can be found on the left; Challenger on the right. Together with a primary conversion goal (`Will users donate to the Wikimedia Foundation?'), this constitutes the mandatory input to start an A/B test using \systemName. The test itself is created for this work, using an old Wikipedia A/B test from 2010 fitted to the current Wikipedia page for \href{https://en.wikipedia.org/wiki/A/B_testing}{A/B testing}. They tested 2 different group photo positions to determine the more effective one in terms of Wikimedia Foundation donations.}
  \Description{The figure shows two side-by-side screenshots of the Wikipedia article “A/B testing,” illustrating two design variants used in an experiment.
Both versions display the same article content, navigation sidebar, and page structure. The primary difference appears at the top of the article body. In the left variant, a banner reads “Please read: An appeal from our staff,” with a compact image and a call-to-action button. In the right variant, the same appeal appears as a full-width banner with a wider group photo.}
  \label{fig:wikipedia-groupphoto-input}
\end{figure*}

Users set up an experiment by providing up to four inputs that inform \systemName of the experiment, along with the corresponding context:

\begin{enumerate}[leftmargin=*]
    \item \textbf{Variant Images} (required): Images of Control and Challenger designs. Following the design principle to \textit{Shift-left in the Design Process} (\Cref{sub:design_implications}), \systemName operates on screenshots rather than fully implemented webpages. This enables teams to validate design directions during ideation and early iteration, when changes are still inexpensive and actionable. For simple experiments, a single viewport for each variant is sufficient. For experiments exceeding a single viewport, users can provide full-page screenshots that agents explore through simulated scrolling, and for multi-page flows (e.g., checkout processes), multiple screenshots can represent various steps (Section~\ref{sec:actions}).
    
    \item \textbf{Conversion Goal} (required): A natural language description of the success metric for the experiment. Goals can range from simple (``button clicks'') to sophisticated (``subscription signups weighted by plan value''). This specification is incorporated into agent prompts to guide evaluation criteria.
    
    \item \textbf{Target Audience} (optional): Constraints on the simulated user distribution, aligning with the principle of \textit{Audience Restrictions} (\Cref{sub:design_implications}). Specifications can include demographics (``age 25--35''), behavioral characteristics (``users arriving from paid search''), or psychographic segments (``price-sensitive comparison shoppers''). When omitted, personas are generated to represent the broad potential audience for the page content.
    
    \item \textbf{Contextual Documents} (optional): Additional documents, such as analytics reports, past experiment results, or domain knowledge, that might contain relevant information for persona generation and simulation. This input, aligns with the principle of \textit{Additional Context} (\Cref{sub:design_implications}). These documents are used for retrieval-augmented generation (RAG)~\citep{lewis2020retrieval} to retrieve relevant information (Section~\ref{sec:rag}).
\end{enumerate}

An example Control image, Challenger image, and primary conversion goal can be found in \Cref{fig:wikipedia-groupphoto-input}.

\subsubsection{Retrieval-Augmented Generation} \label{sec:rag}

When users provide contextual documents, \systemName employs a RAG pipeline to extract relevant information for persona generation and simulation. The pipeline handles two document types: 
Textual documents are chunked and embedded using an embedding model.
For tabular, we employ a two-stage approach: (1) an LLM generates SQL queries, given the table schema and experiment context, (2) query results are retrieved and summarized into textual insights. These summaries are then embedded alongside textual chunks.

During persona generation and simulation, we retrieve the top-$k$ most relevant chunks using nearest neighbor search based on the embeddings. The query for this search is constructed based on the user's inputs, with generated textual descriptions for both variant images. Optional enhancements include Hypothetical Document Embeddings (HyDE) query expansion~\cite{gao_precise_2023} and cross-encoder re-ranking \cite{nogueira2019passage} to improve retrieval precision.

\subsection{Persona Generation}

Personas represent the simulated users who will evaluate both variants. Each persona is represented with 13 attributes organized into four categories:

\begin{enumerate}[leftmargin=0.5cm]
    \item \textbf{Demographics}: Name, age range, occupation, income level, education, location
    \item \textbf{Psychographics}: Interests, goals, pain points, technical savviness
    \item \textbf{Behavioral}: Online behavior patterns, typical browsing context
    \item \textbf{Task-specific}: Specific tasks the persona would perform on the page, circumstances of their visit (e.g., ``researching before a major purchase,'' or ``arrived from a social media ad'')
\end{enumerate}

\begin{figure*}[t]
  \centering
  \includegraphics[width=0.85\textwidth]{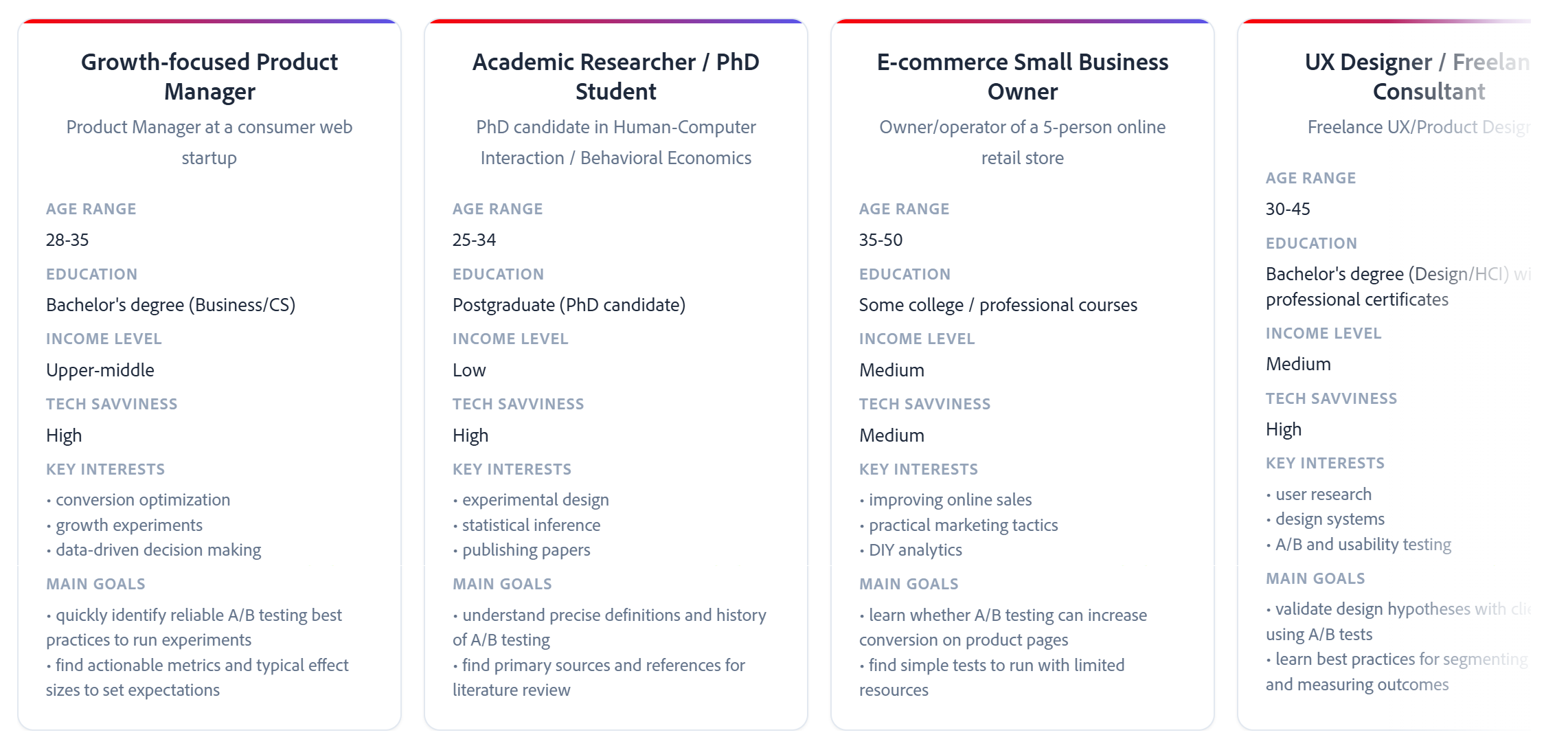}
  \caption{Example Personas. The generated personas include a context, ranging from age, education, and background, to the present context (the main goals in the current browser session).}
  \Description{Four example user personas shown as side-by-side cards. The personas are Growth focused Product Manager, Academic Researcher or PhD Student, E commerce Small Business Owner, and UX Designer or Freelance Consultant. Each card lists age range, education level, income level, tech savviness, key interests, and main goals related to A/B testing and experimentation.}
  \label{fig:wikipedia-groupphoto-personas}
\end{figure*}

Personas are generated by prompting {\small \texttt{GPT-5-mini-2025-08-07}} with the variant screenshots, target audience specification (if provided), and retrieved context from any supplied documents. A sample of generated personas for the Wikimedia donation test can be found in \Cref{fig:wikipedia-groupphoto-personas}. The generation prompt explicitly instructs diversity. Our full prompt can be found in Appendix~\ref{app:persona_prompt}.

\begin{figure}[h]
  \centering
  \includegraphics[width=1\columnwidth]{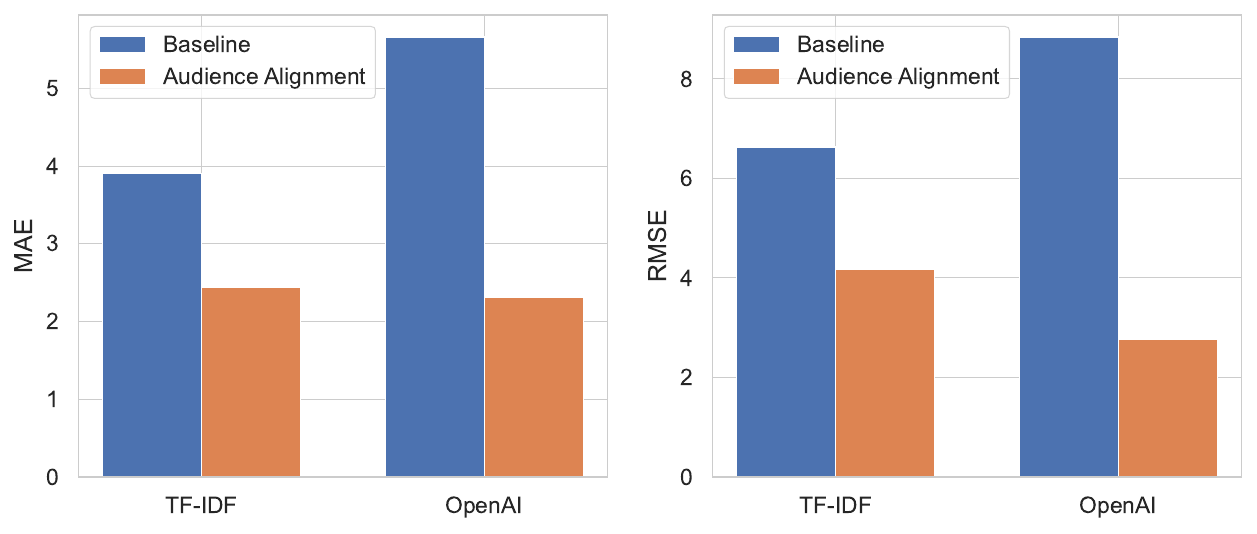}
  \caption{Comparing the target distribution as provided in \textit{audience restrictions} from a user to the distribution generated by the \systemName \textit{persona generation} module. 
  Blue: baseline without providing segment information to the LLM call, orange: using the segments as audience restriction to align generated personas on it.
  Left: absolute mean squared error (MSE), right: root mean squared error (RMSE). 
  Term Frequency-Inverse Document Frequency (TF-IDF): evaluates a word's importance in a document relative to the whole corpus; using cosine-similarity to measure the similarity of words in two collections. Both TF-IDF and direct requests to LLMs were used to classify generated personas to the most similar persona segment (the alignment objective provided in \textit{audience restrictions}).
  Note that outliers (RMSE) in particular are reduced significantly.}
  \Description{The figure presents two bar charts comparing persona generation accuracy with and without audience alignment. Blue bars show a baseline without audience restrictions, and orange bars show results with audience alignment. The left chart reports mean absolute error, and the right chart reports root mean squared error. Results are shown for TF-IDF and OpenAI embeddings. In both metrics and embedding types, audience alignment reduces error, with a larger reduction visible in root mean squared error, indicating fewer large deviations.}
  \label{fig:using_segments_as_audience_restriction}
\end{figure}

\subsubsection{Batched Generation \& Diversity Constraints}

To balance efficiency with diversity, personas are generated in batches of 5--10. Within each batch, previously generated personas are included in the context, encouraging the model to produce distinct individuals. However, across batches, this context is lost, leading to potential duplication. Our evaluation (Section~\ref{sub:persona_diversity_ablation}) quantifies the impact of persona diversity on accuracy.

When users provide known audience segment distributions (e.g., from analytics showing 30\% ``Tech-Savvy Creators,'' 25\% ``Budget-Conscious Browsers''), we align persona generation to match these proportions. Passing in the segment distributions as audience restrictions, the MSE and RMSE are reduced significantly as seen in \Cref{fig:using_segments_as_audience_restriction}. We note that more elaborate schemes to align with arbitrary fixed audience restrictions could be implemented.

\subsection{Persona Simulation}

Each generated persona is executed as an AI agent that evaluates both variants. The evaluation is structured as a multi-turn conversation with the LLM, proceeding through three phases.

\begin{figure*}[t]
  \centering
  \includegraphics[width=0.68\textwidth]{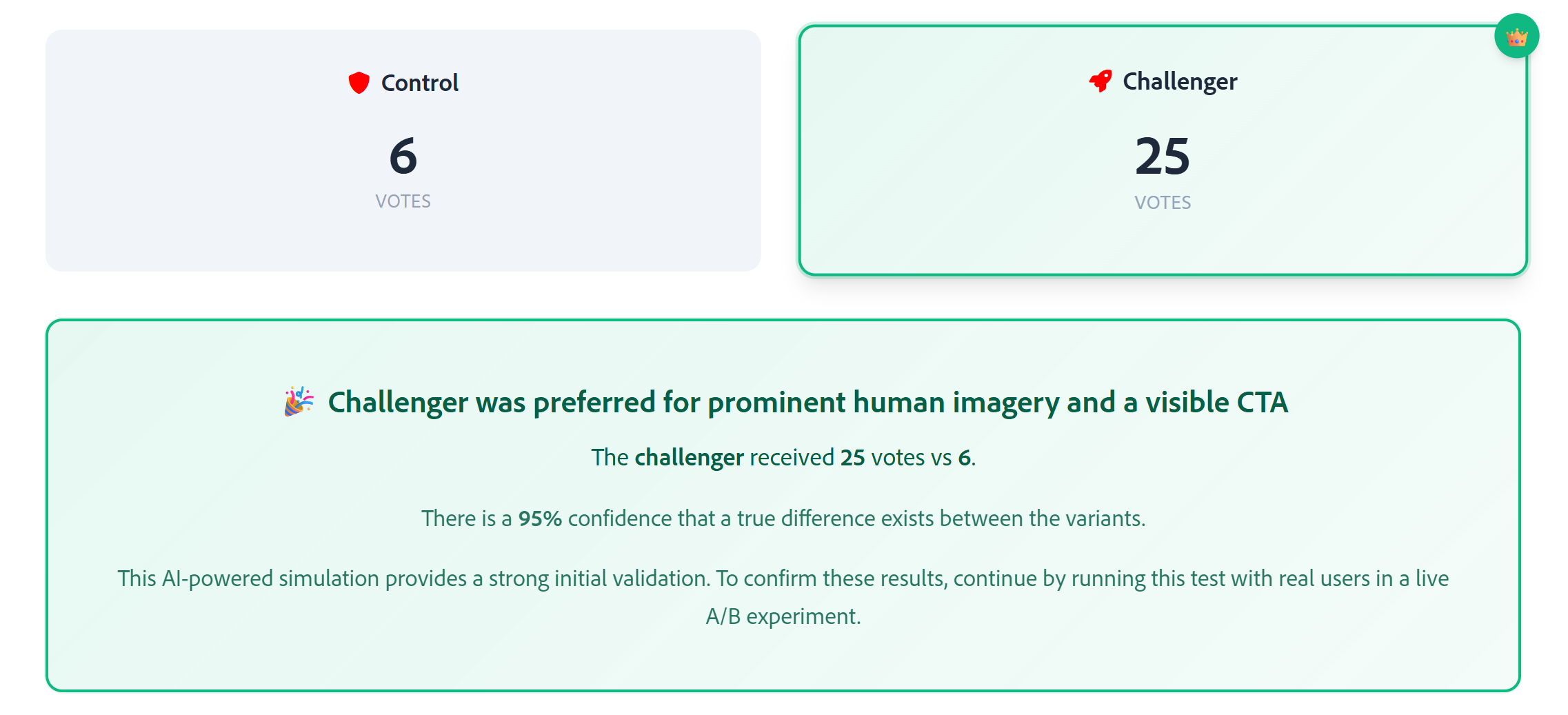}
  \caption{Example Winner. After just 31 agents (1-2 minutes, all started in parallel) conclude, statistical significance is reached. The result coincides with the ground truth data gathered from real users.}
  \Description{A results screen comparing two variants. Control has 6 votes, and Challenger has 25 votes. The interface highlights Challenger as the preferred option, notes a 95\% confidence difference between variants, and states that the AI-based simulation suggests validation before confirmatory testing with real users.}
  \label{fig:wikipedia-groupphoto-winner}
\end{figure*}

The winner of the Wikipedia A/B testing webpage donation banner can be found in \Cref{fig:wikipedia-groupphoto-winner}.

\subsubsection{Contextualization}

The agent receives a system prompt containing:
\begin{enumerate}[leftmargin=0.5cm]
    \item The complete persona specification
    \item The conversion goal, framed as the criterion for preference.
    \item Retrieved context from any provided documents.
    \item Evaluation guidelines covering clarity, friction points, trust signals, value communication, and legal/accessibility compliance.
\end{enumerate}

\subsubsection{Variant Presentation with Counterbalancing} \label{sec:counterbalancing}

The agent receives both variant images in a single prompt, labeled with neutral names (``Image 1'' and ``Image 2''). To mitigate position bias~\cite{zheng_judging_2023}, we employ deterministic counterbalancing by alternating the order in which the variants are presented in between agents.
We further validated the effectiveness of both neutral naming and counterbalancing in Section~\ref{sub:bias_mitigation} to validate that these methods are successful at mitigating biases in the LLM.

\subsubsection{Structured Response}

The agent produces a structured JSON response containing:
\begin{itemize}[leftmargin=0.5cm]
    \item \textbf{Verdict}: One of ``Image 1,'' ``Image 2,'' or ``None''
    \item \textbf{Rationale}: A short explanation identifying specific elements that influenced the preference, in line with the design principle to provide \textit{Rationales \& Actionable Insights} (\Cref{sub:design_implications}).
\end{itemize}

The ``None'' option allows agents to indicate that the persona would not convert on either variant, which serves as an important signal for pages where baseline conversion rates are low.

\subsubsection{Interactive Actions} \label{sec:actions}

For pages that extend beyond the initial viewport or involve multi-step flows, \systemName supports interactive exploration. Agents can invoke tool functions to manipulate their view before making a verdict.
These actions can either be scrolling actions moving the viewport (with configurable aspect ratio) on the provided (full-size) screenshot of a design or custom actions transitioning between different pages of the design, e.g., a checkout button navigating the user to the next page of the checkout flow. 
This capability is critical for testing pages where key conversion elements (e.g., call-to-action elements, pricing) appear below the fold. Without scrolling, agents might evaluate based only on hero content, missing factors that drive real-world conversions.
Agents iterate through actions up to a configurable maximum, stopping early if they express sufficient confidence to make a verdict. 

\subsubsection{Statistical Aggregation} \label{sec:statistics}

After each persona provides their stated preference, we utilize asymptotic confidence sequences \cite{waudby2024time} to test whether the aggregate synthetic preference differs significantly from random chance.
If the synthetic consensus reaches the significance threshold, persona simulation is stopped, and the experiment is completed.
We acknowledge that the use of asymptotic confidence sequences does not meet all necessary assumptions, e.g., due to batched generation of the personas, the samples (personas) are not independent within one batch. Nevertheless, we empirically observe that the approach serves as a highly effective empirical heuristic for early stopping.

\subsection{Summary and Actionable Insight Generation}

\begin{figure*}[t]
  \centering
  \includegraphics[width=0.65\textwidth]{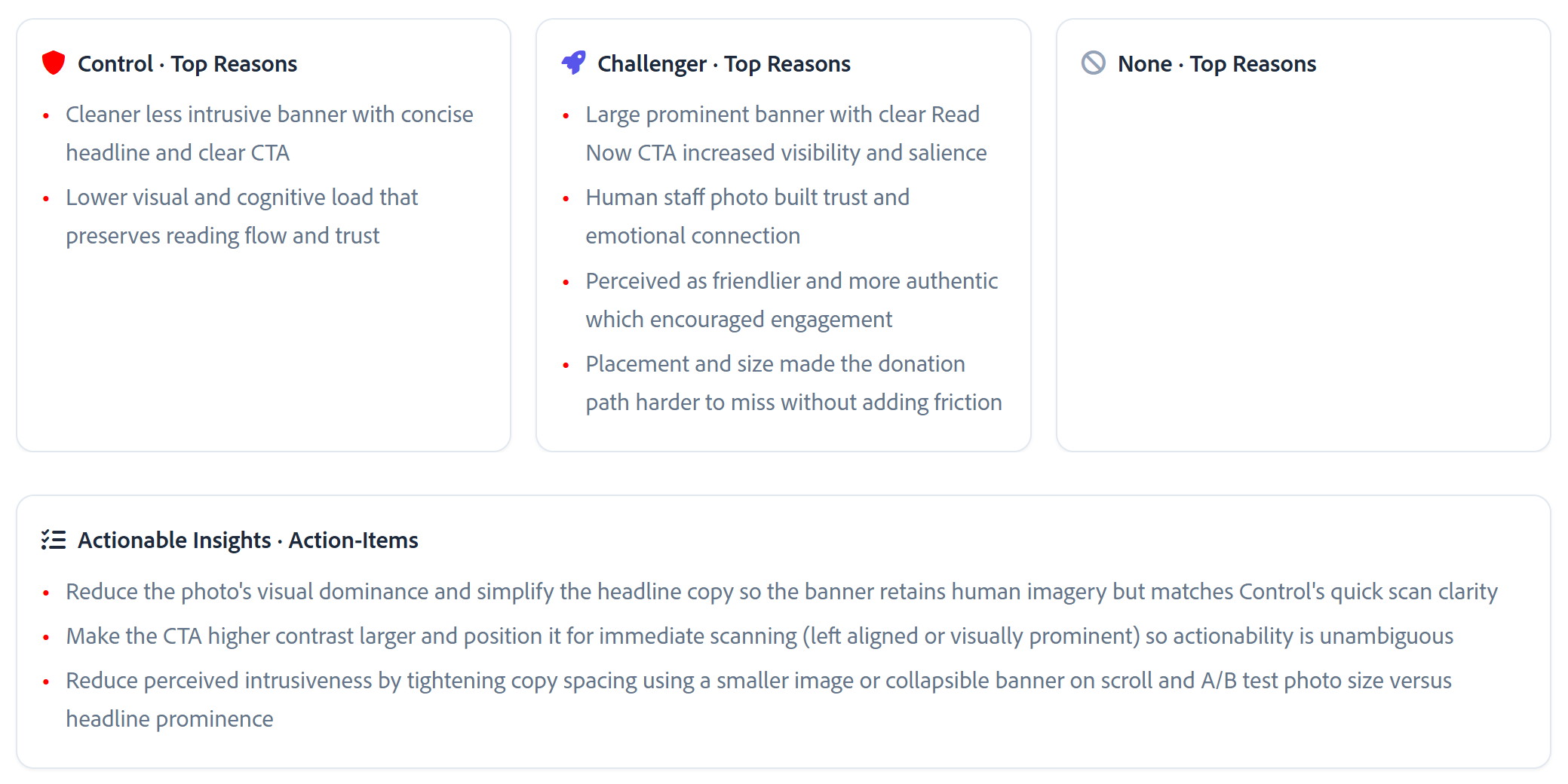}
  \caption{Example Insights. The rationales of all agents are aggregated to show the top reasons why \textit{Control}/\textit{Challenger}/\textit{None} is preferred (for those that preferred \textit{Control}/\textit{Challenger}/\textit{None}). Additionally, \systemName proposes immediate action items to improve and iterate on the Challenger.}
  \Description{An insights dashboard summarizing reasons for preference. Three columns list the top reasons for Control, Challenger, and None, with bullets comparing banner clarity, visual load, human imagery, CTA visibility, and engagement. A section below presents actionable design recommendations to refine the Challenger through adjustments to imagery, headline clarity, CTA contrast, and layout.}
  \label{fig:wikipedia-groupphoto-results}
\end{figure*}

Upon experiment completion, \systemName generates an aggregated summary through a final LLM call. The summary prompt includes all individual rationales and instructs the model to:
\begin{enumerate}[leftmargin=*]
    \item Identify the 3--5 most common themes driving preferences for both Control and Challenger
    \item Surface actionable improvements suggested by dissenting agents
\end{enumerate}
The resulting summary, along with synthetic personas and the statistical verdict, can be exported as a PDF report for stakeholder communication.

The summary and actionable insights for the example Wikipedia A/B test can be found in \Cref{fig:wikipedia-groupphoto-results}.

\subsection{Reproducibility}

We use {\small \texttt{GPT-5-mini-2025-08-07} as the default model for persona generation and {\small \texttt{GPT-5-2025-08-07} as the default model for persona simulation, whilst being model-agnostic subject to requiring multi-modality (image input). We use up to 200 concurrent requests, sufficient to complete experiments in 1--5 minutes with early stopping enabled. We note that a higher degree of concurrency was not required in our use case, but is easily achievable by scaling the number of concurrent agents. Prompt templates can be found in Appendix~\ref{app:prompts}.

\section{Evaluation}
\label{sec:evaluation}

We evaluate the effectiveness of \systemName through five different evaluation measures: 
\begin{enumerate*}
    \item backtesting against historical A/B tests with known outcomes (\Cref{sub:backtesting}),
    \item robustness and consistency evaluation (\Cref{sub:robustness}),
    \item validation of bias mitigation strategies (\Cref{sub:bias_mitigation}),
    \item a persona ablation study (\Cref{sub:persona_ablation}), and
    \item feedback from sessions conducted with multiple different experimentation practitioners (\Cref{sub:deployment_feedback}).
\end{enumerate*}

\subsection{Backtesting Against Historical Tests} \label{sub:backtesting}

To assess if SimAB predictions align with historical experimental outcomes,
we evaluate its accuracy on a corpus of 
47 historical A/B tests with documented outcomes.
The corpus consists of manually curated real-world A/B tests that reached statistical significance, stemming from three sources:
\begin{enumerate*}
    \item large-scale commerce web applications and
    \item a commercial desktop application by our organization
    , and
    \item publicly available tests from the Wikimedia Foundation~\citep{wikimedia2010_test2, wikimedia2011_oct21}.
\end{enumerate*}
The corpus captures a wide range of design changes, including changes to call-to-action elements, headlines, layouts, and imagery, conducted across multiple geographic regions. Further, these experiments targeted varied objectives, such as increasing sign-ups, increasing engagement, or decreasing bounce rates. The data spans a broad time frame, with tests from our organization collected over the last five years and the Wikimedia dataset covering experiments from 2010 and 2011.
For each test, we configured \systemName with the Control and Challenger screenshots, conversion goal, and target audience constraints if available for the test, and ran experiments with an $\alpha=0.05$ to test for statistical significance. \Cref{tab:backtesting} presents the confusion matrix.

\begin{table}[t]
    \centering
    \caption{Confusion matrix \systemName achieved on our corpus of 47 diverse historical A/B tests. Note that two tests did not reach statistical significance and subsequently are not taken into account for the confusion matrix.}
    \label{tab:backtesting}
    \begin{tabular}{ll|cc|c}
        \toprule
        & & \multicolumn{2}{c|}{\textbf{Ground Truth}} & \\
        & & \textbf{Challenger} & \textbf{Control} & \textit{Precision} \\
        \midrule
        \multirow{2}{*}{\textbf{Predicted}} 
         & \textbf{Challenger} & 22 (TP) & 9 (FP) & 71\% \\
         & \textbf{Control}    & 6 (FN)  & 8 (TN) &      \\
        \midrule
         & \textit{Recall}     & 79\%    &        & Acc: 67\% \\
        \bottomrule
    \end{tabular}
\end{table}

The results demonstrate that \systemName provides a strong predictive signal,
agreeing with the historical experimental outcome in two-thirds of cases (67\% agreement rate) and agreeing with 80\% of cases where the historical test favored the challenger (recall).
While this agreement rate, well above the 50\% chance baseline, suggests SimAB provides useful predictive signal,
we note that predictive accuracy represents only one part of the system's overall value. As highlighted by our practitioner feedback (\Cref{sub:deployment_feedback}), the accompanying \emph{rationales} and \emph{actionable insights} were often deemed more valuable than the binary verdict itself. By articulating the reasoning behind a preference, \systemName exposes specific design trade-offs and usability friction, guiding further iteration and refinement even in cases where the statistical prediction is imperfect.

Additionally, we find that including an analytics report that breaks down the performance of webpages over the last three years as a contextual document reduces the number of agents required to reach statistical significance by over 50\%, concluding the experiment significantly quicker, while maintaining the same accuracy.

Beyond the aggregated metrics, we identify a strong correlation between accuracy and the number of agents necessary to reach statistical significance. Stratifying the accuracy by the required number of agents (\Cref{fig:confidence}) reveals that the fewer agents are required to reach statistical significance, the more accurate the results are.
This pattern aligns with intuition: for very distinct variants, fewer agents suffice to accurately detect the preference. When differences are subtle, more agents are needed, and the synthetic signal becomes more uncertain, similar to the cases where real A/B tests also struggle.

\begin{figure}
    \centering
    \includegraphics[width=0.9\linewidth]{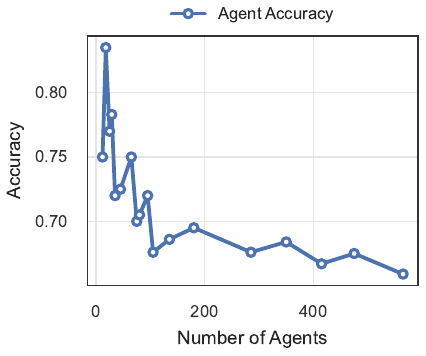}
    \caption{Trade-off between agent count and accuracy. We plot the cumulative accuracy for the subset of experiments that concluded with $\le x$ agents. Experiments that reach statistical significance quickly exhibit substantially higher accuracy compared to those requiring more agents.}
    \Description{A line chart shows agent accuracy versus the number of agents. Accuracy is higher with fewer agents and gradually declines as agent count increases, illustrating a trade-off between speed to significance and overall accuracy.}
    \label{fig:confidence}
\end{figure}

We define \textit{high-confidence} tests as tests for which a significant result is reached within at most 70 agents. 
Filtering by \textit{high-confidence} results, we report 75\% accuracy, 75\% precision, 88\% recall, and 81\% F1-score. 60\% of all tests are \textit{high-confidence} tests.
We note that \textit{very-high-confidence} tests that reach significance within 20 agents exceed 80\% accuracy and 90\% F1-score. However, \textit{very-high-confidence} tests come at the cost of being at a traditional A/B testing market-usual percentage of tests that reach significance of 30\%.

This suggests a practical operating mode where users act on fast-con\-verg\-ing predictions
while treating slower, lower-confidence results as inconclusive to boost accuracy in the significant results (even restricted to \textit{high-confidence} tests, we report 60\% of tests reaching significance as compared to industry estimates of only 20-30\% for traditional A/B testing~\cite{Sharma2025ABTesting}). 
In addition to being more accurate, stopping early increases speed significantly, directly addressing the design requirement to ``prioritize speed over perfect accuracy'' identified in \Cref{sub:design_implications}, enabling the critical, high-velocity feedback loops necessary for time-sensitive design environments. We frequently observed a <1 minute result for those 60\% of fast tests, which is satisfactory for end-user applications and will become more tangible with LLM calls expected to speed up.

\subsection{Robustness and Consistency} \label{sub:robustness}

For \systemName to be practically useful, results must be reproducible and consistent. We conducted three experiments for robustness analyses.

\subsubsection{Repeated Experiments}

\begin{figure}
    \centering
    \includegraphics[width=0.8\linewidth]{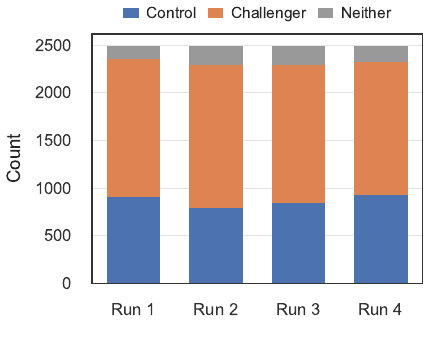}
    \caption{Consistency analysis across four independent runs of the same experiment. While exact vote counts fluctuate slightly due to LLM stochasticity, the system consistently identifies the same winner (Challenger) with a significant margin.}
    \Description{A stacked bar chart compares vote counts for Control, Challenger, and Neither across four runs. Challenger receives the highest number of votes in every run, showing consistent winner identification despite minor variation in counts.}
    \label{fig:robustness_same_experiment}
\end{figure}

To assess stability across runs, we executed the same experiment (a well-differentiated pricing page test) four times.
As shown in \Cref{fig:robustness_same_experiment}, all four runs identified the same winner (Challenger) despite variation in vote distributions. The Challenger preference rate ranged from 56\% to 62\% across runs, which is within expected statistical variation. This demonstrates robustness to the inherent stochasticity in LLM sampling.

\subsubsection{Multi-Challenger Transitivity}

Since \systemName conducts pairwise comparisons, multi-variant tests require running all pairwise comparisons. A critical requirement is \emph{transitivity}: if A beats B and B beats C, then A should beat C. Intransitive results (cycles) would undermine confidence in rankings.

We analyzed 50+ pairwise comparisons from multi-variant experiments curated from the same sources as our backtesting corpus (\Cref{sub:backtesting}). In \emph{all cases}, results were perfectly transitive, establishing a unique total ordering for every design set. We observed zero instances of cyclic preferences (e.g., $A > B > C > A$).
This consistency confirms that \systemName is suitable for iterative ``hill-climbing'' workflows, where designers progressively refine a Challenger against previous winners, and further validates its robustness against random (seed) variations.

\subsubsection{Persona Consistency}

To evaluate if personas provide consistent answers, we tested whether individual personas provide deterministic verdicts when presented with identical inputs. We executed 100 distinct personas 20 times each against the same Control-Challenger pair and measured the \emph{Minority Vote Count}: the number of times an agent deviated from its personas' majority verdict.

\begin{figure}
    \centering
    \includegraphics[width=0.85\linewidth]{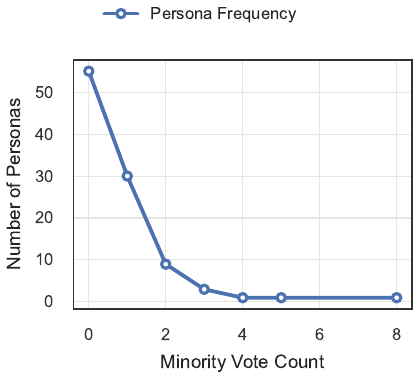}
    \caption{Persona Consistency Analysis. We ran each persona 20 times on identical inputs. The x-axis shows the number of deviation votes (minority votes). The steep drop-off indicates that individual personas hold stable, deterministic preferences, with the majority showing 0 or 1 deviation out of 20 runs.}
    \Description{A line chart shows persona frequency by minority vote count. Most personas show zero or one minority vote, with a sharp drop as deviation increases, indicating stable and consistent preferences.}
    \label{fig:persona_consistency}
\end{figure}

The results, visualized in \Cref{fig:persona_consistency}, demonstrate high consistency. Over 50\% of personas exhibited perfect stability (0 minority votes), identifying the same preference in 20/20 trials. Furthermore, 85\% of personas had two or fewer deviations.
This confirms that \systemName  agents produce internally consistent stated preferences.

\subsection{Bias Mitigation Validation} \label{sub:bias_mitigation}

LLM evaluation biases are well-documented~\cite{zheng_judging_2023}. In this section, we assess whether our approach effectively mitigates these biases.

\subsubsection{Counterbalancing Effectiveness} \label{sub:counterbalancing_eval}

We validated counterbalancing in our system by running an experiment where Control and Challenger are \emph{identical images}. Results after 4,756 agents:
\begin{itemize}[leftmargin=0.5cm]
    \item Votes for image presented first: 2,187
    \item Votes for image presented second: 2,191
    \item ``None'' votes: 378
\end{itemize}

The 4-vote difference is statistically negligible ($\chi^2 = 0.0004$, $p = 0.984$). For comparison, without counterbalancing (Control first), we observed
Challenger being preferred 6--17 times more than Control,
i.e., a bias that would substantially distort results. In \Cref{tab:bias_mitigation_results} we report the effects of combining counterbalancing and neutral naming.

\begin{table}[]
    \caption{Validation for our bias mitigation techniques counterbalancing and neutral names. The table displays the number of votes both variants received during a test conducted with identical images to isolate bias.}
    \label{tab:bias_mitigation_results}
    \centering
    \begin{tabular}{cc|cccc}
        \toprule
        \makecell[b]{\textbf{Counter-}\\ \textbf{balancing}} & 
        \makecell[b]{\textbf{Neutral}\\ \textbf{Names}} &
        \textbf{Control} & \textbf{Challenger} & \textbf{None} & \textbf{p-value}\\  
        \midrule
        \xmark & \xmark & 126 & 749 & 125 & 0.0000 \\
        \xmark & \cmark &  52 & 887 &  61 & 0.0000 \\
        \cmark & \xmark & 442 & 434 & 124 & 0.8131 \\
        \cmark & \cmark & 460 & 469 &  71 & 0.7930 \\
        \bottomrule
    \end{tabular}
\end{table}

\subsection{Persona Ablation Study} \label{sub:persona_ablation}

\subsubsection{Persona Diversity Ablation} 
\label{sub:persona_diversity_ablation}

To assess whether the use of personas impacts accuracy, we compared three setups:

\begin{itemize}[leftmargin=0.5cm]
    \item \textbf{Baseline}: Standard batched generation (5--10 per batch) with no explicit duplication across batches
    \item \textbf{No Diversity}: A single, fixed persona for all agents
    \item \textbf{High Diversity}: De-duplication using embedding\footnote{Embeddings were generated using OpenAI text embeddings based on the textual description of the personas.} cosine similarity ($\theta = 0.85$) with regeneration until diversity was reached. Regeneration was done using persona-to-persona generation~\cite{chan_scaling_2024}
\end{itemize}

\begin{figure}
    \centering
    \includegraphics[width=0.83\linewidth]{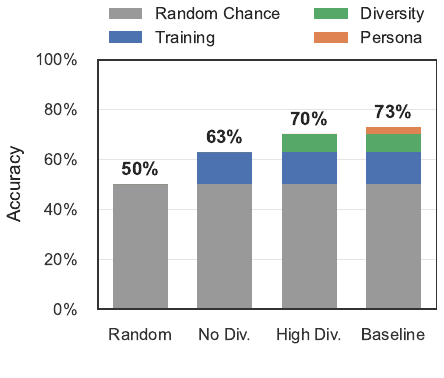}
    \caption{Persona Diversity Ablation study. We show the effect of not using personas at all (that is, a single fixed persona) versus multiple adaptive personas per experiment, subsequently increasing accuracy.}
    \Description{A bar chart compares accuracy across random choice, no diversity, high diversity, and baseline conditions. Accuracy increases when multiple diverse personas are used, outperforming random and single-persona setups.}
    \label{fig:persona_ablation_study}
\end{figure}

The results are summarized in \Cref{fig:persona_ablation_study}. We observe that 
removing diversity degrades agreement rates by 10 percentage points, suggesting that diverse synthetic perspectives improve predictive alignment with real experimental outcomes.
However, extreme diversity (High Diversity condition) did not improve over baseline, suggesting diminishing returns. The baseline approach, which naturally produces moderate diversity through batched generation, appears to strike an effective balance between diversity and efficiency.

\subsubsection{Personas Analyzed per Group}

\begin{figure}[t]
  \centering
  \includegraphics[width=0.95\linewidth]{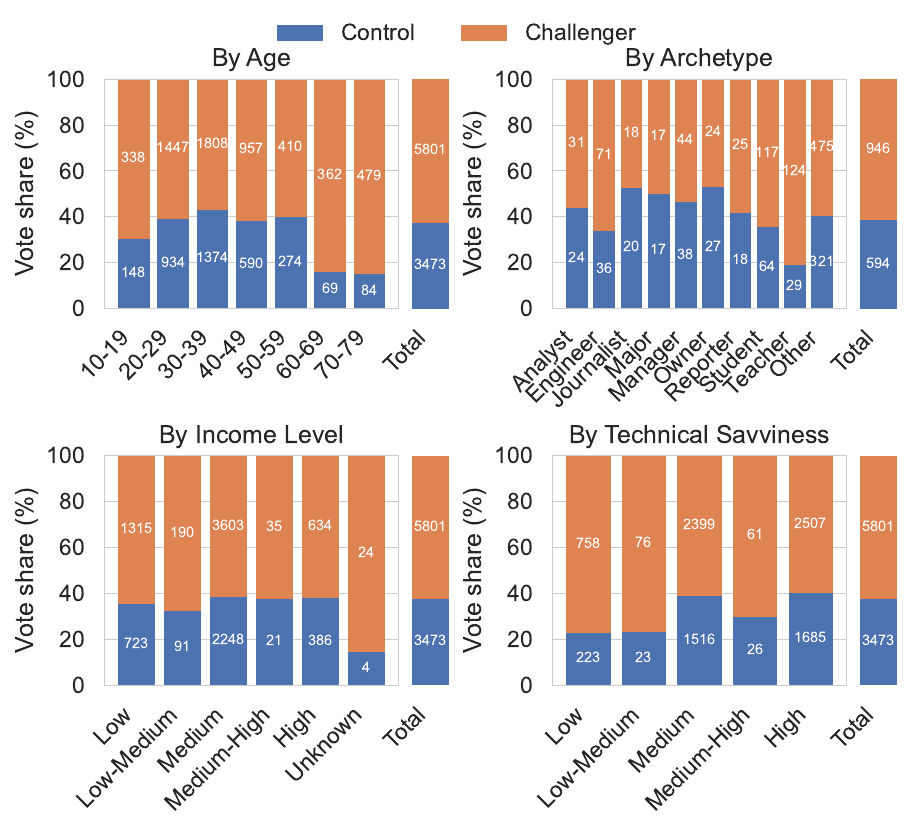}
  \caption{Voting distributions by different persona splits: By age, profession archetype, income level, and technical savviness. Vote share for control is given in blue vs. vote share for challenger in orange by group.}
  \Description{Four stacked bar charts show vote share for Control and Challenger by age, archetype, income level, and technical savviness. Challenger receives a higher vote share across most groups, although different segments of generated personas vote significantly different.}
  \label{fig:voting_behavior_by_groups}
\end{figure}

In the following, we aim to demonstrate that different segments (if split by some classification) of the personas admit different voting behavior. In \Cref{fig:voting_behavior_by_groups}, splits by age, profession archetype, income level, and technical savviness are given. For three of the four splits, at least one of the groups are significantly out of distribution, with a p-value for the age-group `{\small \texttt{\texttt{60-69}}}' of $4\cdot 10^{-32}$, for the archetype-group `{\small \texttt{Teacher}}' of $3\cdot 10^{-7}$ and for the technical-savviness-group `{\small \texttt{\texttt{Low}}}' of $7\cdot 10^{-23}$ to stem from the same distribution as the other groups, demonstrating that different personas behave differently.

In addition, \systemName is sufficient to produce significantly different personas when provided with different input images, even without specifying audience restrictions,
which suggests that persona generation is responsive to input images.

\subsection{Feedback from Practitioners} \label{sub:deployment_feedback}

As part of ongoing participatory design engagements, we evaluated \systemName with the same 14 practitioners from four enterprise organizations described in \Cref{sec:formative}. We maintained this cohort to directly validate whether the system effectively addressed the specific pain points identified in our initial sessions \Cref{sec:formative}.

The evaluation methodology was adapted to fit organizational constraints. Due to organizational policies, external users utilized a ``concierge'' evaluation model: they provided assets for active design questions, and we executed the simulations and shared the results with them. In contrast, four internal practitioners were able to access the \systemName UI directly to run independent experiments. Following these evaluations, we conducted review sessions to analyze the results.

\textbf{Enabling Previously Untestable Scenarios}:
A recurring theme across organizations was the value of \systemName for scenarios that are impractical or infeasible with traditional A/B testing. Participants highlighted low-traffic pages (e.g., niche, regional, or B2B-facing content), as well as changes expected to yield small but meaningful improvements, which directly contributes to the \textit{Low-traffic Pages} pain point identified in \Cref{sub:customer_pain_points}.

Practitioners were excited when they realized that \systemName can be used for pages that do not receive enough traffic to be tested, i.e., that it can not only help with the existing tests, but also enables \textit{completely new pages} to be optimized in a more consistent and data-driven way.

\textbf{Consistency and Alignment Across Stakeholders}:
Several participants noted that \systemName has the potential to help reduce inconsistencies in internal decision-making. Design reviews often involve multiple stakeholders with differing opinions and heuristics, leading to subjective debates. The simulated results and accompanying rationales provided a shared, data-driven reference point that supported alignment and more efficient discussions, validating the design principle to provide \textit{Rationales \& Actionable Insights} (\Cref{sub:design_implications}).

\textbf{Acceptance of Accuracy Trade-offs}:
Participants demonstrated a pragmatic understanding of the system's limitations. Rather than expecting perfect accuracy, teams emphasized transparency, known failure modes, and appropriate use, viewing \systemName as a tool for guiding early decisions and prioritization while reserving live testing for high-stakes candidates. This adoption pattern confirms the validity of our design principles to \textit{Prioritize Speed over Perfect Accuracy} and \textit{Shift-left in the Design Process} (\Cref{sub:design_implications}), as practitioners successfully acknowledged the tool as a rapid filter rather than a replacement for real testing.

\section{Discussion}
\label{sec:discussion}

Our evaluation demonstrates that synthetic audiences can provide a meaningful signal for A/B testing, with an accuracy that varies predictably with test difficulty. In this section, we discuss the implications of these findings, characterize appropriate use cases, acknowledge limitations, and outline directions for future research.

\subsection{Accuracy Results in the Context of Design Research}

Some of the accuracy rates reported may initially seem modest compared to the certainty practitioners expect from traditional A/B testing. However, several contextual factors suggest this performance is more useful than it appears:

\textbf{Stratified Confidence as a Filter}:
The 75\% accuracy on \textit{high-confidence} predictions (tests concluding in $\leq$70 agents as defined in \Cref{sub:backtesting}) and $>$80\% accuracy on \textit{very-high-confidence} predictions ($\leq$20 agents) are substantially more actionable. In practice, users can treat \systemName as a high-precision filter: results that emerge quickly are likely reliable; results requiring many agents are uncertain and warrant live testing. This framing, i.e., that fast results are confident, slow results are uncertain, aligns with intuition about A/B testing generally.

\textbf{Simulation as an Alternative to Intuition}: Traditional A/B tests themselves have non-trivial error rates due to multiple testing, seasonality, and external confounds. The relevant comparison is \systemName versus \emph{no testing at all}, which is the default for low-traffic pages, multi-variant comparisons, and micro-optimizations.
A 67\% agreement rate achieved in 5 minutes may outperform intuition-based decisions, though direct comparison would require further study.

\textbf{The Value of Explanatory Rationales}: Even when verdicts are incorrect, the generated rationales often surface legitimate design considerations. Several practitioners noted using rationales to identify potential issues, even when they disagreed with the overall verdict.

\subsection{Use Cases}

Based on our evaluation and deployment experience, we identify four primary use cases where \systemName provides the most substantial value:

\textbf{Preflight Screening}: The clearest use case is rapid screening of design variants before committing to live experiments. \systemName can quickly eliminate obviously suboptimal options, reducing the variant space for more expensive live testing. One deployment partner described this as ``killing bad ideas fast''. If \systemName strongly rejects a variant, it can be assumed that this simulated A/B test result will likely align with future real users' feedback, saving time and resources for design, marketing,  and product practitioners.

\textbf{Testing the ``Untestable''}: Many design optimization opportunities are structurally excluded from traditional testing due to low-traffic pages, highly specific audience segments, new market exploration, and large variant spaces, e.g., 5 potential designs.
\systemName provides directional signal where previously there was often only intuition.
The 67\% accuracy is most valuable precisely in these scenarios where the alternative is guesswork.

\textbf{Privacy-First Testing}: Unlike traditional A/B testing, \systemName requires no user data collection. This provides a privacy-preserving alternative for organizations with strict data governance requirements (e.g., GDPR compliance) or for testing in privacy-sensitive contexts (e.g., healthcare, finance). The ability to simulate user feedback without observing real user behavior represents a meaningful privacy improvement.

\textbf{Rapid Design Iteration}:
Traditional A/B testing provides feedback only after development is complete. \systemName enables evaluation of mockups and prototypes before any implementation, shifting feedback left in the design process. This changes the iteration cadence from ``\textit{design $\rightarrow$ develop $\rightarrow$ test (months) $\rightarrow$ learn}'' to ``\textit{design $\rightarrow$ test (minutes) $\rightarrow$ refine $\rightarrow$ develop}''.

\subsection{Limitations and Boundary Conditions}

We identify several important limitations that constrain \systemName's applicability:

\textbf{Behavioral Complexity Beyond Screenshots.}
\systemName evaluates static screenshots with optional simulated scrolling, so for tests where these factors are primary drivers, live experiments remain necessary, as it \textit{does not capture}:
\begin{itemize}[leftmargin=0.5cm]
    \item \textit{Performance Effects}: Page load time, responsiveness, and perceived speed significantly impact conversion but are invisible in screenshots.
    \item \textit{Interaction Dynamics}: Hover states, animations, micro-interactions, and complex form behavior influence user experience in ways static images cannot convey.
    \item \textit{Temporal Effects}: Repeat visits, brand perception over time, and network effects unfold across sessions.
\end{itemize}

\textbf{Novel Interaction Patterns}: LLM evaluations reflect patterns learned from training data. For genuinely novel interfaces, e.g., designs that differ substantially from existing web conventions, agents may not accurately predict human responses. The further a design deviates from the training data distribution, the less reliable the predictions become.
We note that such a shift in training distribution versus test-time design might stem from both changes in domain and shifts in human behavior over time.

\textbf{Subtle Effect Sizes}: Our accuracy stratification reveals that \systemName performs best on tests with clear differentiation. For subtle optimizations where even humans struggle to express preferences (i.e., the ``hard cases'' that real A/B tests also struggle with), synthetic evaluation may not provide a reliable signal. This is not unique to our approach; it reflects a fundamental limit of preference elicitation for near-identical options. However, it must be noted that with the increased run time of our simulation (2-10 minutes), 95\% of our tests achieved significance, which is much higher than the typical 30\% for traditional A/B testing \cite{Sharma2025ABTesting}.

\textbf{Representation Validity}: Synthetic personas inherit biases from underlying language models. Santurkar et al.~\cite{santurkar_whose_2023} showed that LLM outputs disproportionately reflect perspectives of Western, educated populations. Our personas may not accurately represent global, cross-cultural, or marginalized user groups. Practitioners should critically evaluate whether generated personas adequately cover their intended audience, particularly for products serving diverse populations.

\textbf{Automation of Manipulation}: A/B testing has long raised concerns about ``dark patterns'', i.e., designs that exploit psychological biases to drive conversions against user interests~\cite{gray_dark_2018}. \systemName's speed and cost make it easier to rapidly iterate on persuasion techniques. Although the tool itself is neutral, responsible deployment requires ethical guidelines on what optimizations are appropriate.

\textbf{Generalization to Other Domains}: Our A/B testing corpus consists primarily of marketing and e-commerce webpages. Further testing and development might be needed to generalize both the results and potentially adapt the system to other domains.

\subsection{Future Directions}

Our work opens up several directions for future research.

\textbf{Domain-Adaptive Fine-Tuning}.
The 67\% accuracy reflects zero-shot performance without domain adaptation. Fine-tuning on historical A/B test outcomes within a specific domain could substantially improve accuracy in that domain. Organizations with extensive experimentation archives could train domain-specific evaluators calibrated to their users.

\textbf{Multi-Modal and Dynamic Evaluation}.
Extending beyond screenshots to evaluate video recordings, interactive prototypes, or even live websites would capture behavioral dynamics currently missing. Integration with tools like Browser Use~\cite{browser_use} for full page interaction could bridge the gap between static and dynamic evaluation.

\textbf{Calibrated Uncertainty Quantification}.
Our current confidence estimates are heuristic. Developing principled uncertainty quantification (calibrated probability estimates for verdict correctness) would enable users to make better decisions.

\textbf{Evaluation of LLM-Facing Interfaces}.
As AI agents increasingly interact with websites (for shopping, information retrieval, etc.), optimizing for LLM ``users'' may become as important as optimizing for humans. \systemName's architecture naturally extends to this emerging use case, evaluating how well designs support AI interaction patterns.

\section{Conclusion}

Traditional A/B testing depends on real user traffic, which creates a fundamental bottleneck limiting experimentation capacity across the web. In this paper, we presented \systemName, a system that reframes A/B testing as a fast, privacy-preserving simulation using persona-conditioned AI agents.
Through formative work with experimentation practitioners, we identified three scenarios where traffic constraints prevent testing: low-traffic pages, multi-variant comparisons, and micro-optimizations. These insights informed a system design centered on speed, actionable rationales, and flexible audience specification.
Our technical contribution comprises a complete pipeline for synthetic A/B testing: diverse persona generation from webpage content, evaluation by counterbalanced AI agents, sequential statistical aggregation enabling valid early stopping, and rationale synthesis for actionable insights. We demonstrated that counterbalancing effectively neutralizes position bias (4-vote difference across 4,756 identical-image evaluations, $p > 0.99$) and that persona diversity improves accuracy by 10 percentage points.
Evaluation against 47 historical A/B tests with known outcomes revealed 67\% overall alignment, improving to 75\% and $>$80\% for \textit{high-confidence} and \textit{very-high-confidence} predictions that conclude quickly. Perfect transitivity across 50+ multi-variant comparisons confirms consistent rankings. Deployment with four enterprise partners validated practical utility: teams valued rapid screening, explanatory rationales, and the ability to test previously infeasible scenarios.

We position \systemName not as a replacement for live experimentation, but as a complementary tool that expands the boundaries of what can be tested. The stratified accuracy results suggest a practical operating mode:
treat fast-converging results as stronger directional signals; treat slow-converging results as weak hypotheses warranting live validation.
This ``preflight check'' framing, i.e., eliminating bad ideas fast, screening the variant space, and shifting feedback left in the design process, captures the primary value proposition.
We believe \systemName represents a \textit{shift in who can conduct rigorous design evaluation}. Traditional A/B testing requires substantial traffic, technical infrastructure, and organizational patience. Synthetic A/B testing democratizes experimentation, making data-driven design accessible to small organizations, emerging markets, and individual designers.

As generative AI transforms both content creation and user interaction patterns, the need for rapid evaluation will only grow.
SimAB demonstrates that synthetic audiences can provide predictive signal that aligns with historical experimental outcomes more often than chance.
We hope this work catalyzes further research on AI-assisted evaluation and contributes to a future where data-driven design is accessible to all.


\begin{acks}
We thank David Arbour for valuable discussions and feedback on this work.
\end{acks}

\bibliographystyle{ACM-Reference-Format}
\bibliography{references}

@inproceedings{ding_designgpt_2023,
	title = {{DesignGPT}: {Multi}-{Agent} {Collaboration} in {Design}},
	shorttitle = {{DesignGPT}},
	url = {https://ieeexplore.ieee.org/document/10494260},
	doi = {10.1109/ISCID59865.2023.00056},
	urldate = {2025-09-05},
	booktitle = {2023 16th {International} {Symposium} on {Computational} {Intelligence} and {Design} ({ISCID})},
	author = {Ding, Shiying and Chen, Xinyi and Fang, Yan and Liu, Wenrui and Qiu, Yiwu and Chai, Chunlei},
	month = dec,
	year = {2023},
	note = {ISSN: 2473-3547},
	pages = {204--208},
}

@inproceedings{he_webvoyager_2024,
	address = {Bangkok, Thailand},
	title = {{WebVoyager}: {Building} an {End}-to-{End} {Web} {Agent} with {Large} {Multimodal} {Models}},
	shorttitle = {{WebVoyager}},
	url = {https://aclanthology.org/2024.acl-long.371/},
	doi = {10.18653/v1/2024.acl-long.371},
	urldate = {2025-09-04},
	booktitle = {Proceedings of the 62nd {Annual} {Meeting} of the {Association} for {Computational} {Linguistics} ({Volume} 1: {Long} {Papers})},
	publisher = {Association for Computational Linguistics},
	author = {He, Hongliang and Yao, Wenlin and Ma, Kaixin and Yu, Wenhao and Dai, Yong and Zhang, Hongming and Lan, Zhenzhong and Yu, Dong},
	editor = {Ku, Lun-Wei and Martins, Andre and Srikumar, Vivek},
	month = aug,
	year = {2024},
	pages = {6864--6890},
}

@inproceedings{zheng_evalignux_2025,
	address = {New York, NY, USA},
	series = {{CHI} '25},
	title = {{EvAlignUX}: {Advancing} {UX} {Evaluation} through {LLM}-{Supported} {Metrics} {Exploration}},
	isbn = {979-8-4007-1394-1},
	shorttitle = {{EvAlignUX}},
	url = {https://dl.acm.org/doi/10.1145/3706598.3714045},
	doi = {10.1145/3706598.3714045},
	urldate = {2025-07-14},
	booktitle = {Proceedings of the 2025 {CHI} {Conference} on {Human} {Factors} in {Computing} {Systems}},
	publisher = {Association for Computing Machinery},
	author = {Zheng, Qingxiao and Chen, Minrui and Sharma, Pranav and Tang, Yiliu and Oswal, Mehul and Liu, Yiren and Huang, Yun},
	month = apr,
	year = {2025},
	pages = {1--25},
}

@inproceedings{park_generative_2023,
	address = {San Francisco CA USA},
	title = {Generative {Agents}: {Interactive} {Simulacra} of {Human} {Behavior}},
	isbn = {979-8-4007-0132-0},
	shorttitle = {Generative {Agents}},
	url = {https://dl.acm.org/doi/10.1145/3586183.3606763},
	doi = {10.1145/3586183.3606763},
	language = {en},
	urldate = {2025-06-23},
	booktitle = {Proceedings of the 36th {Annual} {ACM} {Symposium} on {User} {Interface} {Software} and {Technology}},
	publisher = {ACM},
	author = {Park, Joon Sung and O'Brien, Joseph and Cai, Carrie Jun and Morris, Meredith Ringel and Liang, Percy and Bernstein, Michael S.},
	month = oct,
	year = {2023},
	pages = {1--22},
}

@misc{chen_towards_2025,
	title = {Towards a {Design} {Guideline} for {RPA} {Evaluation}: {A} {Survey} of {Large} {Language} {Model}-{Based} {Role}-{Playing} {Agents}},
	shorttitle = {Towards a {Design} {Guideline} for {RPA} {Evaluation}},
	url = {http://arxiv.org/abs/2502.13012},
	doi = {10.48550/arXiv.2502.13012},
	urldate = {2025-05-19},
	publisher = {arXiv},
	author = {Chen, Chaoran and Yao, Bingsheng and Zou, Ruishi and Hua, Wenyue and Lyu, Weimin and Ye, Yanfang and Li, Toby Jia-Jun and Wang, Dakuo},
	month = mar,
	year = {2025},
	note = {arXiv:2502.13012 [cs]},
}

@misc{agent_ab_2024,
	title = {AgentA/B: Automated and Scalable Web A/B Testing with Interactive LLM Agents},
	author = {Junzhe Chen and others},
	year = {2024},
	publisher = {arXiv},
	url = {https://arxiv.org/abs/2400.00000},
}

@article{kohavi_online_2013,
	title = {Online Controlled Experiments at Large Scale},
	author = {Kohavi, Ron and Deng, Alex and Frasca, Brian and Walker, Toby and Xu, Ya and Pohlmann, Nils},
	journal = {Proceedings of the 19th ACM SIGKDD International Conference on Knowledge Discovery and Data Mining},
	year = {2013},
	doi = {10.1145/2487575.2488217},
}

@book{slivkins_introduction_2019,
	title = {Introduction to Multi-Armed Bandits},
	author = {Slivkins, Aleksandrs},
	year = {2019},
	publisher = {Foundations and Trends in Machine Learning},
	doi = {10.1561/2200000068},
}

@article{argyle_out_2023,
	title = {Out of One, Many: Using Language Models to Simulate Human Samples},
	author = {Argyle, Lisa P. and Busby, Ethan C. and Fulda, Nancy and Gubler, Joshua R. and Rytting, Christopher and Wingate, David},
	journal = {Political Analysis},
	volume = {31},
	number = {3},
	pages = {337--351},
	year = {2023},
	doi = {10.1017/pan.2023.2},
}

@article{horton_large_2023,
	title = {Large Language Models as Simulated Economic Agents: What Can We Learn from Homo Silicus?},
	author = {Horton, John J.},
	journal = {NBER Working Paper},
	year = {2023},
	doi = {10.3386/w31122},
}

@inproceedings{santurkar_whose_2023,
	title = {Whose Opinions Do Language Models Reflect?},
	author = {Santurkar, Shibani and Durmus, Esin and Ladhak, Faisal and Lee, Cinoo and Liang, Percy and Hashimoto, Tatsunori},
	booktitle = {Proceedings of the 40th International Conference on Machine Learning},
	year = {2023},
}

@inproceedings{zheng_judging_2023,
	title = {Judging LLM-as-a-Judge with MT-Bench and Chatbot Arena},
	author = {Zheng, Lianmin and Chiang, Wei-Lin and Sheng, Ying and Zhuang, Siyuan and Wu, Zhanghao and Zhuang, Yonghao and Lin, Zi and Li, Zhuohan and Li, Dacheng and Xing, Eric P. and Zhang, Hao and Gonzalez, Joseph E. and Stoica, Ion},
	booktitle = {Advances in Neural Information Processing Systems},
	year = {2023},
}

@inproceedings{zhou_webarena_2024,
	title = {WebArena: A Realistic Web Environment for Building Autonomous Agents},
	author = {Zhou, Shuyan and Xu, Frank F. and Zhu, Hao and Zhou, Xuhui and Lo, Robert and Sridhar, Abishek and Cheng, Xianyi and Bisk, Yonatan and Fried, Daniel and Alon, Uri and Neubig, Graham},
	booktitle = {International Conference on Learning Representations},
	year = {2024},
}

@misc{browser_use,
	title = {Browser Use: AI Browser Automation},
	url = {https://github.com/browser-use/browser-use},
	year = {2024},
	author = {Browser Use},
}

@inproceedings{gray_dark_2018,
	title = {The Dark (Patterns) Side of UX Design},
	author = {Gray, Colin M. and Kou, Yubo and Battles, Bryan and Hoggatt, Joseph and Toombs, Austin L.},
	booktitle = {Proceedings of the 2018 CHI Conference on Human Factors in Computing Systems},
	year = {2018},
	doi = {10.1145/3173574.3174108},
}

@misc{gao_precise_2023,
	title = {Precise Zero-Shot Dense Retrieval without Relevance Labels},
	author = {Gao, Luyu and Ma, Xueguang and Lin, Jimmy and Callan, Jamie},
	year = {2023},
	publisher = {arXiv},
	doi = {10.48550/arXiv.2212.10496},
}

@misc{chan_scaling_2024,
	title = {Scaling Synthetic Data Creation with 1,000,000,000 Personas},
	author = {Chan, Xin and Wang, Xiaoyang and Yu, Dian and Mi, Haitao and Yu, Dong},
	year = {2024},
	publisher = {arXiv},
	doi = {10.48550/arXiv.2406.20094},
}

@article{waudby2024time,
  title={Time-uniform central limit theory and asymptotic confidence sequences},
  author={Waudby-Smith, Ian and Arbour, David and Sinha, Ritwik and Kennedy, Edward H and Ramdas, Aaditya},
  journal={The Annals of Statistics},
  volume={52},
  number={6},
  pages={2613--2640},
  year={2024},
  publisher={Institute of Mathematical Statistics}
}

@misc{wang_agentab_2025,
	title = {{AgentA/B}: Automated and Scalable Web {A/B} Testing with Interactive {LLM} Agents},
	author = {Wang, Dakuo and Hsu, Ting-Yao and Lu, Yuxuan and Gu, Hansu and Cui, Limeng and Xie, Yaochen and Headean, William and Yao, Bingsheng and Veeragouni, Akash and Liu, Jiapeng and Nag, Sreyashi and Wang, Jessie},
	year = {2025},
	eprint = {2504.09723},
	archiveprefix = {arXiv},
	primaryclass = {cs.HC},
	doi = {10.48550/arXiv.2504.09723},
	url = {http://arxiv.org/abs/2504.09723},
}

@inproceedings{lu_uxagent_2025,
	title = {{UXAgent}: An {LLM} Agent-Based Usability Testing Framework for Web Design},
	author = {Lu, Yuxuan and Yao, Bingsheng and Gu, Hansu and Huang, Jing and Wang, Jessie and Li, Laurence and Gesi, Jiri and He, Qi and Li, Toby Jia-Jun and Wang, Dakuo},
	booktitle = {CHI Extended Abstracts},
	year = {2025},
	doi = {https://doi.org/10.1145/3706599.3719729},
}

@article{park_generative_agents_1000,
	title = {Generative Agent Simulations of 1,000 People},
	author = {Park, Joon Sung and Zou, Carolyn Q and Shaw, Aaron and Hill, Benjamin Mako and Cai, Carrie and Morris, Meredith Ringel and Willer, Robb and Liang, Percy and Bernstein, Michael S},
	year = {2024},
	journal = {arXiv preprint arXiv:2411.10109},
}

@article{sarstedt_silicon_2024,
	title = {Using Large Language Models as Silicon Samples: Practical Guidelines and Best Practices},
	author = {Sarstedt, Marko and Adler, Susanne J. and Ringle, Christian M.},
	journal = {Psychological Methods},
	year = {2024},
	doi = {https://doi.org/10.1002/mar.21982},
}

@article{toubia_twin2k500_2025,
	title = {Database Report: {Twin-2K-500}: A Data Set for Building Digital Twins of over 2,000 People Based on Their Answers to over 500 Questions},
	author = {Toubia, Olivier and Gui, George Z. and Peng, Tianyi and Merlau, Daniel J. and Li, Ang and Chen, Haozhe},
	journal = {Marketing Science},
	volume = {44},
	number = {6},
	pages = {1446--1455},
	year = {2025},
	doi = {10.1287/mksc.2025.0262},
}

@misc{peng_megastudy_2025,
	title = {A Mega-Study of Digital Twins Reveals Strengths, Weaknesses and Opportunities for Further Improvement},
	author = {Peng, Tianyi and Gui, George and Merlau, Daniel J. and Fan, Grace Jiarui and Sliman, Malek Ben and Brucks, Melanie and Johnson, Eric J. and Morwitz, Vicki and others},
	year = {2025},
	eprint = {2509.19088},
	archiveprefix = {arXiv},
	primaryclass = {cs.AI},
	doi = {10.48550/arXiv.2509.19088},
}

@misc{hu_population_aligned_2025,
	title = {Population-Aligned Persona Generation for {LLM}-based Social Simulation},
	author = {Hu, Zhengyu and Xiao, Zheyuan and Xiong, Max and Lei, Yuxuan and Wang, Tianfu and Lian, Jianxun and Ding, Kaize and Xiao, Ziang and Yuan, Nicholas Jing and Xie, Xing},
	year = {2025},
	journal = {arXiv preprint arXiv:2502.09499},
}

@inproceedings{samuel_personagym_2024,
	title = {{PersonaGym}: Evaluating Persona Agents and {LLMs}},
	author = {Samuel, Vinay and Zou, Henry Peng and Zhou, Yue and Chaudhari, Shreyas and Kalyan, A. and Rajpurohit, Tanmay and Deshpande, A. and Narasimhan, Karthik and Murahari, Vishvak},
	booktitle = {Conference on Empirical Methods in Natural Language Processing},
	year = {2024},
}

@inproceedings{zhu_reliable_simulator_2024,
	title = {How Reliable is Your Simulator? Analysis on the Limitations of Current {LLM}-based User Simulators for Conversational Recommendation},
	author = {Zhu, Lixi and Huang, Xiaowen and Sang, Jitao},
	booktitle = {Companion Proceedings of the Web Conference 2024},
	year = {2024},
	doi = {10.1145/3589335.3651489},
}

@article{agarwal_synthetic_ab,
	title = {Synthetic {A/B} Testing Using Synthetic Interventions},
	author = {Agarwal, Anish and Shah, Devavrat and Shen, Dennis},
	journal = {arXiv preprint},
	year = {2024},
}

@misc{kolluri_socrates_2025,
	title = {Finetuning {LLMs} for Human Behavior Prediction in Social Science Experiments},
	author = {Kolluri, Akaash and Wu, Shengguang and Park, Joon Sung and Bernstein, Michael S.},
	year = {2025},
	eprint = {2509.05830},
	archiveprefix = {arXiv},
	doi = {10.48550/arXiv.2509.05830},
}

@article{gu_synthetic_users_2025,
	title = {Synthetic Users: Insights from Designers' Interactions with Persona-Based Chatbots},
	author = {Gu, (Eric) Heng and Chandrasegaran, Senthil and Lloyd, Peter},
	journal = {Artificial Intelligence for Engineering Design, Analysis and Manufacturing},
	volume = {39},
	pages = {e2},
	year = {2025},
	doi = {10.1017/S0890060424000283},
}

@article{schmidt_simulating_hcd_2024,
	title = {Simulating the Human in {HCD} with {ChatGPT}: Redesigning Interaction Design with {AI}},
	author = {Schmidt, Albrecht and Elagroudy, Passant and Draxler, Fiona and Kreuter, Frauke and Welsch, Robin},
	journal = {Interactions},
	volume = {31},
	number = {1},
	pages = {24--31},
	year = {2024},
	doi = {10.1145/3637436},
}

@article{wang_user_behavior_2025,
	title = {User Behavior Simulation with Large Language Model-based Agents},
	author = {Wang, Lei and Zhang, Jingsen and Yang, Hao and Chen, Zhi-Yuan and Tang, Jiakai and Zhang, Zeyu and Chen, Xu and Lin, Yankai and Sun, Hao and Song, Ruihua and Zhao, Xin and Xu, Jun and Dou, Zhicheng and Wang, Jun and Wen, Ji-Rong},
	journal = {ACM Transactions on Information Systems},
	volume = {43},
	number = {2},
	pages = {55:1--55:37},
	year = {2025},
	doi = {10.1145/3708985},
}

@misc{wang_opera_2025,
	title = {{OPeRA}: A Dataset of Observation, Persona, Rationale, and Action for Evaluating {LLMs} on Human Online Shopping Behavior Simulation},
	author = {Wang, Ziyi and Lu, Yuxuan and Li, Wenbo and Amini, Amirali and Sun, Bo and Bart, Yakov and Lyu, Weimin and Gesi, Jiri and Wang, Tian and Huang, Jing and Su, Yu and Ehsan, Upol and Alikhani, Malihe and Li, Toby Jia-Jun and Chilton, Lydia and Wang, Dakuo},
	year = {2025},
	journal = {arXiv preprint arXiv:2502.07974},
}

@inproceedings{mansour_paars_2025,
	title = {{PAARS}: Persona Aligned Agentic Retail Shoppers},
	author = {Mansour, Saab and Perelli, Leonardo and Mainetti, Lorenzo and Davidson, George and D'Amato, Stefano},
	booktitle = {Proceedings of the 1st Workshop for Research on Agent Language Models (REALM 2025)},
	year = {2025},
}

@misc{sun_llm_agent_agentic_2025,
	title = {{LLM} Agent Meets Agentic {AI}: Can {LLM} Agents Simulate Customers to Evaluate Agentic-{AI}-based Shopping Assistants?},
	author = {Sun, Lu and Fu, Shihan and Yao, Bingsheng and Lu, Yuxuan and Li, Wenbo and Gu, Hansu and Gesi, Jiri and Huang, Jing and Luo, Chen and Wang, Dakuo},
	year = {2025},
	eprint = {2509.21501},
	archiveprefix = {arXiv},
	doi = {10.48550/arXiv.2509.21501},
}

@misc{amin_genai_personas_2025,
	title = {How Is Generative {AI} Used for Persona Development?: A Systematic Review of 52 Research Articles},
	author = {Amin, Danial and Salminen, Joni and Ahmed, Farhan and Tervola, Sonja M. H. and Sethi, Sankalp and Jansen, Bernard J.},
	year = {2025},
	eprint = {2504.04927},
	archiveprefix = {arXiv},
	doi = {10.48550/arXiv.2504.04927},
}

@misc{pourasad_genai_usability_2024,
	title = {Does {GenAI} Make Usability Testing Obsolete?},
	author = {Pourasad, Ali Ebrahimi and Maalej, Walid},
	year = {2024},
	eprint = {2411.00634},
	archiveprefix = {arXiv},
	doi = {10.48550/arXiv.2411.00634},
}

@inproceedings{brand_gpt_market_2024,
	title = {Using {GPT} for Market Research},
	author = {Brand, James and Israeli, A. and Ngwe, Donald},
	booktitle = {ACM Conference on Economics and Computation},
	year = {2024},
}

@misc{lu_multiturn_2025,
	title = {Can {LLM} Agents Simulate Multi-Turn Human Behavior? Evidence from Real Online Customer Behavior Data},
	author = {Lu, Yuxuan and Huang, Jing and Han, Yan and Bei, Bennet and Xie, Yaochen and Wang, Dakuo and Wang, Jessie and He, Qi},
	year = {2025},
}

@misc{piao_agentsociety_2025,
	title = {{AgentSociety}: Large-Scale Simulation of {LLM}-Driven Generative Agents Advances Understanding of Human Behaviors and Society},
	author = {Piao, J. and Yan, Yuwei and Zhang, Jun and Li, Nian and Yan, Junbo and Lan, Xiaochong and Lu, Zhihong and Zheng, Zhiheng and Wang, Jing Yi and Zhou, Di and Gao, Chen and Xu, Fengli and Zhang, Fang and Rong, Ke and Su, Jun and Li, Yong},
	year = {2025},
	journal = {arXiv preprint},
}

@article{tan_user_modeling_2023,
	title = {User Modeling in the Era of Large Language Models: Current Research and Future Directions},
	author = {Tan, Zhaoxuan and Jiang, Meng},
	journal = {IEEE Data Engineering Bulletin},
	year = {2023},
}

@inproceedings{zhang_agentcf_2023,
	title = {{AgentCF}: Collaborative Learning with Autonomous Language Agents for Recommender Systems},
	author = {Zhang, Junjie and Hou, Yupeng and Xie, Ruobing and Sun, Wenqi and McAuley, Julian and Zhao, Wayne Xin and Lin, Leyu and Wen, Ji-rong},
	booktitle = {The Web Conference},
	year = {2023},
}

@inproceedings{zhang_generative_agents_rec_2023,
	title = {On Generative Agents in Recommendation},
	author = {Zhang, An and Sheng, Leheng and Chen, Yuxin and Li, Hao and Deng, Yang and Wang, Xiang and Chua, Tat-Seng},
	booktitle = {Annual International ACM SIGIR Conference on Research and Development in Information Retrieval},
	year = {2023},
}

@inproceedings{chen_recusersim_2025,
	title = {{RecUserSim}: A Realistic and Diverse User Simulator for Evaluating Conversational Recommender Systems},
	author = {Chen, Luyu and Dai, Quanyu and Zhang, Zeyu and Feng, Xueyang and Zhang, Mingyu and Tang, Pengcheng and Chen, Xu and Zhu, Yue and Dong, Zhenhua},
	booktitle = {The Web Conference},
	year = {2025},
}

@misc{wang_customerr1_2025,
	title = {{Customer-R1}: Personalized Simulation of Human Behaviors via {RL}-based {LLM} Agent in Online Shopping},
	author = {Wang, Ziyi and Lu, Yuxuan and Zhang, Yimeng and Huang, Jing and Wang, Dakuo},
	year = {2025},
	journal = {arXiv preprint},
}

@inproceedings{shao_characterllm_2023,
	title = {{Character-LLM}: A Trainable Agent for Role-Playing},
	author = {Shao, Yunfan and Li, Linyang and Dai, Junqi and Qiu, Xipeng},
	booktitle = {Conference on Empirical Methods in Natural Language Processing},
	year = {2023},
}

@misc{buck_blueprint_2025,
	title = {{BluePrint}: A Social Media User Dataset for {LLM} Persona Evaluation and Training},
	author = {Bück-Kaeffer, Aurélien and Chooi, Je Qin and Zhao, Dan and Touzel, M. P. and Pelrine, Kellin and Godbout, J. and Rabbany, Reihaneh and Yang, Zachary},
	year = {2025},
	journal = {arXiv preprint},
}

@misc{abdulhai_consistent_personas_2025,
	title = {Consistently Simulating Human Personas with Multi-Turn Reinforcement Learning},
	author = {Abdulhai, Marwa and Cheng, Ryan and Clay, Donovan and Althoff, Tim and Levine, Sergey and Jaques, Natasha},
	year = {2025},
	journal = {arXiv preprint},
}

@inproceedings{wang_human_vs_agent_2025,
	title = {Human vs. Agent in Task-Oriented Conversations},
	author = {Wang, Zhefan and Geng, Ning and Guo, Zhiqiang and Ma, Weizhi and Zhang, Min},
	booktitle = {Proceedings of the 2025 Annual International ACM SIGIR Conference on Research and Development in Information Retrieval in the Asia Pacific Region},
	year = {2025},
}

@article{zhao_thompson_ctr_2025,
	title = {Optimizing Click-Through Rates in Online Advertising Using Thompson Sampling},
	author = {Zhao, Ben},
	journal = {ITM Web of Conferences},
	volume = {73},
	pages = {01012},
	year = {2025},
	doi = {10.1051/itmconf/20257301012},
}

@inproceedings{gopalan_thompson_2014,
	title = {Thompson Sampling for Complex Online Problems},
	author = {Gopalan, Aditya and Mannor, Shie and Mansour, Yishay},
	booktitle = {Proceedings of the 31st International Conference on Machine Learning},
	series = {Proceedings of Machine Learning Research},
	volume = {32},
	number = {1},
	pages = {100--108},
	year = {2014},
	publisher = {PMLR},
}

@inproceedings{duanUICritEnhancingAutomated2024,
	title = {{{UICrit}}: {{Enhancing Automated Design Evaluation}} with a {{UI Critique Dataset}}},
	shorttitle = {{{UICrit}}},
	booktitle = {Proceedings of the 37th {{Annual ACM Symposium}} on {{User Interface Software}} and {{Technology}}},
	author = {Duan, Peitong and Cheng, Chin-Yi and Li, Gang and Hartmann, Bjoern and Li, Yang},
	date = {2024-10-11},
	series = {{{UIST}} '24},
	pages = {1--17},
	publisher = {Association for Computing Machinery},
	location = {New York, NY, USA},
	doi = {10.1145/3654777.3676381},
	url = {https://doi.org/10.1145/3654777.3676381},
	urldate = {2026-01-09},
	isbn = {979-8-4007-0628-8},
}

@inproceedings{duanGeneratingUIDesign2023,
	title = {Towards {{Generating UI Design Feedback}} with {{LLMs}}},
	booktitle = {Adjunct {{Proceedings}} of the 36th {{Annual ACM Symposium}} on {{User Interface Software}} and {{Technology}}},
	author = {Duan, Peitong and Warner, Jeremy and Hartmann, Bjoern},
	year = {2023},
	month = oct,
	series = {{{UIST}} '23 {{Adjunct}}},
	pages = {1--3},
	publisher = {Association for Computing Machinery},
	address = {New York, NY, USA},
	doi = {10.1145/3586182.3615810},
	urldate = {2026-01-09},
	isbn = {979-8-4007-0096-5},
}

@inproceedings{duanGeneratingAutomaticFeedback2024,
	title = {Generating {{Automatic Feedback}} on {{UI Mockups}} with {{Large Language Models}}},
	booktitle = {Proceedings of the 2024 {{CHI Conference}} on {{Human Factors}} in {{Computing Systems}}},
	author = {Duan, Peitong and Warner, Jeremy and Li, Yang and Hartmann, Bjoern},
	date = {2024-05-11},
	series = {{{CHI}} '24},
	pages = {1--20},
	publisher = {Association for Computing Machinery},
	location = {New York, NY, USA},
	doi = {10.1145/3613904.3642782},
	url = {https://doi.org/10.1145/3613904.3642782},
	urldate = {2026-01-09},
	isbn = {979-8-4007-0330-0},
}

@article{duanVisualPromptingIterative2025,
	title = {Visual {{Prompting}} with {{Iterative Refinement}} for {{Design Critique Generation}}},
	author = {Duan, Peitong and Cheng, Chin-Yi and Hartmann, Bjoern and Li, Yang},
	date = {2025-05-22},
	eprint = {2412.16829},
	eprinttype = {arXiv},
	eprintclass = {cs},
	doi = {10.48550/arXiv.2412.16829},
	url = {http://arxiv.org/abs/2412.16829},
	urldate = {2026-01-09},
	pubstate = {prepublished},
}

@misc{Sharma2025ABTesting,
  author       = {Sharma, Disha},
  title        = {A/B Testing Statistical Significance: How and When to End a Test},
  year         = {2025},
  month        = feb,
  howpublished = {\url{https://www.convert.com/blog/a-b-testing/experiments-statistical-significance-speed/}},
  note         = {Accessed: 13 January 2026},
}

@article{Kaufmann2016BestArm,
  title={On the complexity of best-arm identification in multi-armed bandit models},
  author={Kaufmann, Emilie and Cappé, Olivier and Garivier, Aurélien},
  journal={Journal of Machine Learning Research},
  year={2016},
  volume={17},
  pages={1--42}
}

@inproceedings{Jamieson2014BestArmSurvey,
  title={Best-arm identification algorithms for multi-armed bandits in the fixed confidence setting},
  author={Jamieson, Kevin and Nowak, Robert},
  booktitle={Information Sciences and Systems},
  year={2014}
}

@inproceedings{benharrakWriterDefinedAIPersonas2024,
	title = {Writer-{{Defined AI Personas}} for {{On-Demand Feedback Generation}}},
	booktitle = {Proceedings of the 2024 {{CHI Conference}} on {{Human Factors}} in {{Computing Systems}}},
	author = {Benharrak, Karim and Zindulka, Tim and Lehmann, Florian and Heuer, Hendrik and Buschek, Daniel},
	date = {2024-05-11},
	series = {{{CHI}} '24},
	pages = {1--18},
	publisher = {Association for Computing Machinery},
	location = {New York, NY, USA},
	doi = {10.1145/3613904.3642406},
	url = {https://doi.org/10.1145/3613904.3642406},
	urldate = {2026-01-09},
	isbn = {979-8-4007-0330-0},
}

@online{choiProxonaSupportingCreators2025,
	title = {Proxona: {{Supporting Creators}}' {{Sensemaking}} and {{Ideation}} with {{LLM-Powered Audience Personas}}},
	shorttitle = {Proxona},
	author = {Choi, Yoonseo and Kang, Eun Jeong and Choi, Seulgi and Lee, Min Kyung and Kim, Juho},
	date = {2025-02-19},
	eprint = {2408.10937},
	eprinttype = {arXiv},
	eprintclass = {cs},
	doi = {10.48550/arXiv.2408.10937},
	url = {http://arxiv.org/abs/2408.10937},
	urldate = {2026-01-09},
	pubstate = {prepublished},
}

@inproceedings{haCloChatUnderstandingHow2024,
	title = {{{CloChat}}: {{Understanding How People Customize}}, {{Interact}}, and {{Experience Personas}} in {{Large Language Models}}},
	shorttitle = {{{CloChat}}},
	booktitle = {Proceedings of the 2024 {{CHI Conference}} on {{Human Factors}} in {{Computing Systems}}},
	author = {Ha, Juhye and Jeon, Hyeon and Han, Daeun and Seo, Jinwook and Oh, Changhoon},
	date = {2024-05-11},
	series = {{{CHI}} '24},
	pages = {1--24},
	publisher = {Association for Computing Machinery},
	location = {New York, NY, USA},
	doi = {10.1145/3613904.3642472},
	url = {https://doi.org/10.1145/3613904.3642472},
	urldate = {2026-01-09},
	isbn = {979-8-4007-0330-0},
}

@inproceedings{hamalainenEvaluatingLargeLanguage2023,
	title = {Evaluating {{Large Language Models}} in {{Generating Synthetic HCI Research Data}}: A {{Case Study}}},
	shorttitle = {Evaluating {{Large Language Models}} in {{Generating Synthetic HCI Research Data}}},
	booktitle = {Proceedings of the 2023 {{CHI Conference}} on {{Human Factors}} in {{Computing Systems}}},
	author = {Hämäläinen, Perttu and Tavast, Mikke and Kunnari, Anton},
	date = {2023-04-19},
	series = {{{CHI}} '23},
	pages = {1--19},
	publisher = {Association for Computing Machinery},
	location = {New York, NY, USA},
	doi = {10.1145/3544548.3580688},
	url = {https://doi.org/10.1145/3544548.3580688},
	urldate = {2026-01-09},
	isbn = {978-1-4503-9421-5},
}

@inproceedings{pangUnderstandingLLMificationCHI2025,
	title = {Understanding the {{LLM-ification}} of {{CHI}}: {{Unpacking}} the {{Impact}} of {{LLMs}} at {{CHI}} through a {{Systematic Literature Review}}},
	shorttitle = {Understanding the {{LLM-ification}} of {{CHI}}},
	booktitle = {Proceedings of the 2025 {{CHI Conference}} on {{Human Factors}} in {{Computing Systems}}},
	author = {Pang, Rock Yuren and Schroeder, Hope and Smith, Kynnedy Simone and Barocas, Solon and Xiao, Ziang and Tseng, Emily and Bragg, Danielle},
	date = {2025-04-25},
	series = {{{CHI}} '25},
	pages = {1--20},
	publisher = {Association for Computing Machinery},
	location = {New York, NY, USA},
	doi = {10.1145/3706598.3713726},
	url = {https://doi.org/10.1145/3706598.3713726},
	urldate = {2026-01-09},
	isbn = {979-8-4007-1394-1},
}

@inproceedings{parkSocialSimulacraCreating2022,
	title = {Social {{Simulacra}}: {{Creating Populated Prototypes}} for {{Social Computing Systems}}},
	shorttitle = {Social {{Simulacra}}},
	booktitle = {Proceedings of the 35th {{Annual ACM Symposium}} on {{User Interface Software}} and {{Technology}}},
	author = {Park, Joon Sung and Popowski, Lindsay and Cai, Carrie and Morris, Meredith Ringel and Liang, Percy and Bernstein, Michael S.},
	date = {2022-10-28},
	series = {{{UIST}} '22},
	pages = {1--18},
	publisher = {Association for Computing Machinery},
	location = {New York, NY, USA},
	doi = {10.1145/3526113.3545616},
	url = {https://doi.org/10.1145/3526113.3545616},
	urldate = {2026-01-09},
	isbn = {978-1-4503-9320-1},
}

@inproceedings{pruittPersonasPracticeTheory2003,
	title = {Personas: Practice and Theory},
	shorttitle = {Personas},
	booktitle = {Proceedings of the 2003 Conference on {{Designing}} for User Experiences},
	author = {Pruitt, John and Grudin, Jonathan},
	date = {2003-06-06},
	series = {{{DUX}} '03},
	pages = {1--15},
	publisher = {Association for Computing Machinery},
	location = {New York, NY, USA},
	doi = {10.1145/997078.997089},
	url = {https://doi.org/10.1145/997078.997089},
	urldate = {2026-01-09},
	isbn = {978-1-58113-728-6},
}

@inproceedings{shinPosterMateAudiencedrivenCollaborative2025,
	title = {{{PosterMate}}: {{Audience-driven Collaborative Persona Agents}} for {{Poster Design}}},
	shorttitle = {{{PosterMate}}},
	booktitle = {Proceedings of the 38th {{Annual ACM Symposium}} on {{User Interface Software}} and {{Technology}}},
	author = {Shin, Donghoon and Lee, Daniel and Hsieh, Gary and Chan, Gromit Yeuk-Yin},
	date = {2025-09-27},
	series = {{{UIST}} '25},
	pages = {1--20},
	publisher = {Association for Computing Machinery},
	location = {New York, NY, USA},
	doi = {10.1145/3746059.3747769},
	url = {https://doi.org/10.1145/3746059.3747769},
	urldate = {2026-01-09},
	isbn = {979-8-4007-2037-6},
}

@inproceedings{suhStoryEnsembleEnablingDynamic2025,
	title = {{{StoryEnsemble}}: {{Enabling Dynamic Exploration}} \& {{Iteration}} in the {{Design Process}} with {{AI}} and {{Forward-Backward Propagation}}},
	shorttitle = {{{StoryEnsemble}}},
	booktitle = {Proceedings of the 38th {{Annual ACM Symposium}} on {{User Interface Software}} and {{Technology}}},
	author = {Suh, Sangho and Lai, Michael and Pu, Kevin and Dow, Steven P. and Grossman, Tovi},
	date = {2025-09-27},
	series = {{{UIST}} '25},
	pages = {1--36},
	publisher = {Association for Computing Machinery},
	location = {New York, NY, USA},
	doi = {10.1145/3746059.3747772},
	url = {https://doi.org/10.1145/3746059.3747772},
	urldate = {2026-01-09},
	isbn = {979-8-4007-2037-6},
}

@online{wangLargeLanguageModels2025,
	title = {Large Language Models That Replace Human Participants Can Harmfully Misportray and Flatten Identity Groups},
	author = {Wang, Angelina and Morgenstern, Jamie and Dickerson, John P.},
	date = {2025-02-03},
	eprint = {2402.01908},
	eprinttype = {arXiv},
	eprintclass = {cs},
	doi = {10.48550/arXiv.2402.01908},
	url = {http://arxiv.org/abs/2402.01908},
	urldate = {2026-01-10},
	pubstate = {prepublished},
}

@article{depaoliUserPersonasIdeation2026,
  title = {User Personas, Ideation and Large Language Models: {{A}} Post-Hoc Study},
  shorttitle = {User Personas, Ideation and Large Language Models},
  author = {De Paoli, Stefano},
  date = {2026-01-01},
  journaltitle = {International Journal of Human-Computer Studies},
  shortjournal = {International Journal of Human-Computer Studies},
  volume = {208},
  pages = {103690},
  issn = {1071-5819},
  doi = {10.1016/j.ijhcs.2025.103690},
  url = {https://www.sciencedirect.com/science/article/pii/S1071581925002472},
  urldate = {2026-01-14}
}

@inproceedings{shinUnderstandingHumanAIWorkflows2024,
  title = {Understanding {{Human-AI Workflows}} for {{Generating Personas}}},
  booktitle = {Proceedings of the 2024 {{ACM Designing Interactive Systems Conference}}},
  author = {Shin, Joongi and Hedderich, Michael A. and Rey, Bartłomiej Jakub and Lucero, Andrés and Oulasvirta, Antti},
  date = {2024-07-01},
  series = {{{DIS}} '24},
  pages = {757--781},
  publisher = {Association for Computing Machinery},
  location = {New York, NY, USA},
  doi = {10.1145/3643834.3660729},
  url = {https://doi.org/10.1145/3643834.3660729},
  urldate = {2026-01-14},
  isbn = {979-8-4007-0583-0}
}

@article{nogueira2019passage,
  title={Passage Re-ranking with BERT},
  author={Nogueira, Rodrigo and Cho, Kyunghyun},
  journal={arXiv preprint arXiv:1901.04085},
  year={2019}
}

@article{lewis2020retrieval,
  title={Retrieval-augmented generation for knowledge-intensive nlp tasks},
  author={Lewis, Patrick and Perez, Ethan and Piktus, Aleksandra and Petroni, Fabio and Karpukhin, Vladimir and Goyal, Naman and K{\"u}ttler, Heinrich and Lewis, Mike and Yih, Wen-tau and Rockt{\"a}schel, Tim and others},
  journal={Advances in neural information processing systems},
  volume={33},
  pages={9459--9474},
  year={2020}
}

@misc{wikimedia2010_test2,
  author       = {{Wikimedia Foundation}},
  title        = {Fundraising 2010: Example Test 2},
  year         = {2010},
  howpublished = {\url{https://meta.wikimedia.org/wiki/Fundraising_2010/Report/Example_Test_2}},
  note         = {Accessed: 2026-01-19}
}

@misc{wikimedia2011_oct21,
  author       = {{Wikimedia Foundation}},
  title        = {Fundraising 2011: Test Updates (October 21)},
  year         = {2011},
  howpublished = {\url{https://meta.wikimedia.org/wiki/Fundraising_2011/Test_Updates/October/21}},
  note         = {Accessed: 2026-01-19}
}


\appendix
\onecolumn
\section{Prompts} \label{app:prompts}

This section details the core prompts utilized in \systemName. While our experimental evaluation and prompts focus on e-commerce environments, preliminary observations suggest that these prompts generalize effectively to other domains, such as marketing campaigns or donation banners. However, we exclude these additional domains from our quantitative analysis due to the lack of historical ground truth data in these domains.

\subsection{Persona Generation Prompt Template}
\label{app:persona_prompt}

The following template shows the structure of prompts used for persona generation. Variables in curly braces are populated dynamically for each evaluation.

\begin{lstlisting}[style=prompt]
SYSTEM_PROMPT = """You are an expert in user research and persona development that creates realistic user personas for e-commerce A/B testing.
Generate diverse personas that represent different user types who might visit this website."""

RESTRICTIONS_TEMPLATE = """Additionally, there are some restrictions to the above fields that you HAVE TO follow. **Important**:
{restrictions}
"""

USER_PROMPT_IMAGE_TEMPLATE = """Analyze the website in the image and generate {num_personas} detailed target audience personas that would likely visit this website.

For each persona, provide the following information in JSON format:
- name: A descriptive name for the persona
- age_range: Age range (e.g., "25-35", "18-24", "45-60")
- occupation: Professional role or background
- income_level: Income bracket (e.g., "Low", "Medium", "High", "Upper-middle")
- education: Education level
- location: Geographic location or type
- interests: Key interests and hobbies
- goals: Primary goals and motivations
- pain_points: Challenges and frustrations
- technical_savviness: Technical comfort level (Low/Medium/High)
- online_behavior: How they typically use the internet
- tasks: List of 3-5 specific tasks they would want to perform on this website
- context: Current session context - where the user is coming from and immediate circumstances

{maybe_restrictions}

Generate personas with meaningful diversity across: privacy orientation (strict <-> relaxed), tolerance for initial friction (low <-> high), time sensitivity (rushed <-> leisurely), and comfort with personalization (low <-> high).

- Personas come to the product from different channels (social media, ads, browsing, LLMs, i.e. organic, owned and paid traffic), most are interested in your product, but not all are 100% ready to convert already, i.e. seeing the website will affect their behavior.
- Personas span casual browsing, product research, and purchasing intent; vary age, risk tolerance, decisiveness, and tech literacy.
- Personas might reject ads, cookies, promotions if they are overly aggressive or fishy.
- Do not assume uniform behavior across personas; ensure the cohort reflects the full range of attitudes described above and reflects the distribution of your actual user base.
- Think in a broader context, do not generate personas where all of their painpoints, tasks and goals are only exactly this product or product category, but rather related to this product. Make sure the personas have a life and needs beyond this product.

Focus on realistic personas that would actually visit this type of website. Base your analysis on the actual content provided, 
considering the products, services, messaging, and target audience indicated by the website content.

As additional context, you have the following snippets retrieved from documents that might be relevant to the persona generation:
{context}

Return the response as a valid JSON array of persona objects.
"""
\end{lstlisting}

\subsection{Persona Simulation Prompt Template}
\label{app:evaluation_prompt}

The following template shows the structure of prompts used for persona simulation. Variables in curly braces are populated dynamically for each evaluation.

\begin{lstlisting}[style=prompt]
SYSTEM PROMPT: You are an expert AI that analyzes e-commerce websites to predict conversion potential through the eyes of a given persona.

You will be shown two e-commerce webpage versions and need to predict which one would generate a conversion from your persona, i.e. more revenue for the company. You can also choose that neither version would result in a purchase if the persona wouldn't be interested in buying from either page.

{EvaluationPrompts.CONVERSION_GOAL_TEMPLATE(conversion_goal)}

{EvaluationPrompts.VALUE_HYPOTHESIS(hypothesis)}

You should evaluate the pages from the perspective of a user with these characteristics, which are **YOUR PRIMARY CHARACTERISTICS**:
{persona_string}

The two versions are labeled '{LLMPref.CONTROL}' and '{LLMPref.CHALLENGER}'. The order of presentation is randomized and the labels do NOT imply quality. Base your prediction strictly on what you see.

Personas are browsing, are interested, but don't always need something right away.
- May browse casually and take time to make purchasing decisions.
- Bounce if the website is not appealing enough, considering content aesthetics and user-friendliness.
- If they bounce, they explain why.

Evaluate these e-commerce pages considering BOTH immediate conversion factors AND long-term customer satisfaction:
**Immediate Conversion Factors:**
- Clarity of pricing and next steps
- Ease of completing the purchase
- Minimal friction and confusion

- Clear pricing and shipping information
- Mobile-friendliness and responsive design
- Search and filtering capabilities
- Overall shopping experience and conversion potential

**Long-term Trust & Satisfaction Factors:**
- Transparency about costs and commitments
- Clear cancellation and refund policies
- Legal compliance and trust signals
- Subscription management clarity

**Critical Success Factors (in order of importance):**
1. **Conversion Likelihood**: Will this persona actually complete the purchase?
2. **Cognitive Load**: Does complexity prevent or enable the decision?
3. **Trust vs. Friction**: Does additional detail build necessary trust or create analysis paralysis?
4. **Persona-Specific Needs**: What does THIS specific user actually care about?

NOTE: Revenue/ARR/and the like do ask of the conversion * price. That is, to determine the revenue/ARR, you need to multiply the conversion by the price of the bought product.

**Decision Framework:**
1. What is this persona's PRIMARY concern? (price, features, compliance, speed, etc.)
2. What is their decision-making style? (quick/impulsive vs. thorough/analytical)
3. What would make them abandon the purchase? (complexity, unclear pricing, lack of details)
4. Which version better matches their specific concern and style?
**Conversion Psychology Guidelines:**
- More information is better ONLY if the persona needs it to feel confident
- Simplicity wins when the persona is price-sensitive, time-pressed, or tech-averse
- Complexity wins when the persona is risk-averse, compliance-focused, or making enterprise decisions
- Consider choice paralysis: too many options can reduce conversion
- Consider cognitive load: complex layouts can overwhelm certain personas
- consider in your evaluation the psychological effects of color
**Example Scenarios:**
- Budget-conscious users often prefer clear, simple pricing over detailed feature comparisons
- Time-pressed users may abandon complex decision trees in favor of obvious choices
- Non-technical users can be overwhelmed by detailed feature lists
- Individual buyers may see team features as irrelevant clutter
- Students and startups often prioritize speed and low commitment over comprehensive information

{EvaluationPrompts.RAG_CONTEXT_TEMPLATE(rag_context)}

Consider that a version might convert better if it builds sufficient trust to overcome slightly increased complexity, or if it prevents post-purchase regret and cancellations.
*NOTE*: Remember that you are a human persona, that has a high focus on visual and subjective reasoning and follow the principles of your persona. Still do NOT neglect any other aspect of your shopping experience. Choose the version where a conversion is more likely for this specific user persona, or choose '{LLMPref.NONE}' if this persona would not make a purchase from either version.
**Note** that better for the persona is meant under the assumption that the persona does not have knowledge of the two website versions, but just of the single one. That means, as an example, that the cheaper version is not necessarily the better option for the company, if the user would buy the more expensive one too because it is actually worth its price.
***IMPORTANT***: Some screenshots might contain redacted prices or unknown prices (XX.XX). Replace them with sensible prices for this market. They are not obscuring or untrustworthy, but rather not yet filled in (draft) or redacted. THEY SHOULD NOT CHANGE YOUR OPINION OF THE WEBSITE TO THE BETTER OR TO THE WORSE.

USER PROMPT: Analyze both e-commerce pages carefully, considering whether the given user persona would buy on either page based on product presentation, trustworthiness, ease of shopping, pricing clarity, and overall user experience. Choose the version that would more likely result in a conversion (must be exactly '{LLMPref.CONTROL}' or '{LLMPref.CHALLENGER}') or '{LLMPref.NONE}' if your persona would not buy from either version. Base your reasoning on specific e-commerce factors that would matter to someone with the given characteristics.
*Important*: Predict which version would generate more ACTUAL CONVERSIONS from this specific persona. Consider whether this persona is more likely to abandon due to complexity or due to lack of information. Different personas have different tolerance for detail vs. simplicity.
\end{lstlisting}

\subsection{Summary Generation Prompt Template}

The following template shows the structure of prompts used for summary generation. Variables in curly braces are populated dynamically for each evaluation.

\begin{lstlisting}[style=prompt]
Analyze the following feedback from AI agents who evaluated two e-commerce website versions.
Each agent chose either Control, Challenger, or None (would not buy from either).

Control was chosen {control_count} times, Challenger was chosen {challenger_count} times, None was chosen {none_count} times.

IMPORTANT: The winner is {winner}.

Please identify the main patterns and themes in their decision-making and provide actionable insights to improve the Challenger.

{feedback_text}

Provide:
1. A TINY summary of why agents preferred one version over another. Only THE ONE AND MAIN reason for it, no intro, no outro. One sentence, no commas, no subordinate clause. Eg. Control won because ... or Challenger was preferred ...
2. The top 1-2 reasons (if Control lost) or 3-5 reasons (if Control won) why agents chose Control (if any)
3. The top 1-2 reasons (if Challenger lost) or 3-5 reasons (if Challenger won) why agents chose Challenger (if any)
4. The top 1-2 reasons why agents chose not to buy from either version (if any). Leave empty ([]) if only a minority of agents chose None.
5. The top 1-3 actionable insights for the experiment, if any. Can be empty if no agents chose Control. If some agents chose Control, provide actionable insights to improve the Challenger. That is, what might be improved to make the Challenger more preferred, and even if the Challenger won by a non-overwhelming margin, "What might be improved to make the results more significant?".

Focus on e-commerce factors like: product presentation, trustworthiness, ease of use, pricing clarity, navigation, mobile-friendliness, and conversion optimization.
\end{lstlisting}

\subsection{RAG Prompt Templates}

The following templates show the structure of prompts used for the RAG part of the input processing. Variables in curly braces are populated dynamically for each evaluation.

\begin{lstlisting}[style=prompt]
HYDE_SYSTEM_PROMPT = (
    "You are an expert writer who generates short factual passages "
    "that *could plausibly answer* a given query. "
    "Write a single, coherent paragraph (3-5 sentences) that directly "
    "and confidently answers the query as if you knew the correct answer. "
    "Do not ask questions, list options, or mention uncertainty. "
    "Do not include meta-comments or instructions."
)
HYDE_USER_PROMPT_TEMPLATE = "Query: {query}"

PERSONA_RAG_QUERY_TEMPLATE = "What are personas that might be interested in a website with the following textual description: {description}?"

EXPERIMENT_RAG_QUERY_TEMPLATE = "What information is relevant for evaluating an A/B test with the following details: Conversion goal: {conversion_goal}. Image descriptions for the variants: {image_descriptions}"

SQL_QUERY_SYSTEM_PROMPT = (
    "You are an expert SQL analytics assistant for A/B testing. "
    "Your workflow involves two steps: \n"
    "1. You first write a DuckDB SQL query (using the table named 'df') that surfaces relevant context about user behavior, business metrics, and current trends based on the experiment context and dataset summary provided. Your output for this step should be *only* the SQL query (no comments or surrounding text). \n"
    "2. Once you see the query result, you then write a brief, accessible summary that provides context for personas evaluating the experiment. \n"
    "Guidelines for the query step: \n"
    "- Write clear, efficient SQL queries using DuckDB syntax. "
    "- Focus on aggregations, segmentations, and filters that reveal user behavior patterns, business trends, or relevant context. \n"
    "- Query metrics and dimensions that help understand the current state and why this experiment matters. \n"
    "- Always query FROM df. \n"
    "- Use descriptive column aliases. \n"
    "- Include LIMIT clauses when appropriate for exploratory or large-result queries. \n"
    "- Output *only* the SQL query, no explanation. \n"
    "Guidelines for the summary step: \n"
    "- Write for non-technical personas/users who will evaluate this experiment, NOT for data scientists or experiment designers. \n"
    "- Provide business context, user behavior insights, and current state information. \n"
    '- Focus on "what is happening" rather than "how to design the test". \n'
    '- Use accessible language - avoid statistical jargon like "CV", "covariate adjustment", "power", "sample size", etc. \n'
    "- Be concise (3-5 sentences max) but informative. \n"
    "- Help personas understand why this experiment is relevant given what the data shows. \n"
    "- Do NOT include experiment design recommendations, statistical methods, or technical guidance. \n"
    "- Frame insights as observable facts or trends, not as actionable items for test design. \n\n"
    'Example good summary: "The data shows that weekly subscription revenue has grown significantly over the past year, from near zero to around $60k per week on average. New subscriber counts closely track revenue growth, with recent weeks showing 10% week-over-week fluctuations. Page visit traffic also correlates with subscriptions, suggesting that visitor volume and conversion both matter for growth." \n\n'
    'Example bad summary: "Use Gross New Subs as a primary metric. Account for 66% CV with longer test durations. Consider blocked randomization by week to avoid confounding from upward trend. Use covariate adjustment to reduce variance." \n\n'
    "You will always be prompted first to write a query, and then after seeing the result, to write a summary."
)

SQL_QUERY_USER_PROMPT_TEMPLATE = (
    "Given the following context, write {num_queries} SQL queries to provide insights about the experiment. \n"
    "Experiment context: \n"
    "- Primary Conversion Goal: {conversion_goal} \n"
    "- Hypothesis: {hypothesis} \n"
    "Description of the dataset: \n"
    "{dataset_description}"
    "Audience: \n"
    "{persona_restrictions}"
)
\end{lstlisting}

\section{Example Generated Personas}
\label{app:personas}

Table~\ref{tab:example_personas} shows three example personas generated for a software pricing page evaluation.

\begin{table*}[t]
\centering
\caption{Example personas generated for a creative software pricing page.}
\label{tab:example_personas}
\begin{tblr}{
  colspec = {X[l,0.15] X[l,0.28] X[l,0.28] X[l,0.28]},
  row{1} = {font=\bfseries\small},
  hlines,
  vlines,
  rowsep = 2pt,
}
Attribute & Persona 1 & Persona 2 & Persona 3 \\
Name & ``Creative Director Carlos'' & ``Freelance Designer Priya'' & ``Student Creator Alex'' \\
Age & 38--45 & 28--32 & 19--22 \\
Occupation & Creative Director at ad agency & Freelance graphic designer & University student, design major \\
Income & High (\$120k+) & Medium (\$50--70k) & Low (part-time work) \\
Goals & Team license, reliable support & Cost-effective solo plan & Student discount, essential features \\
Pain Points & Complex procurement process & Unpredictable income, need flexibility & Limited budget, overwhelmed by options \\
Tech Savvy & High & High & Medium \\
Context & Evaluating tools for 15-person team & Comparing before annual renewal & Researching for class project \\
\end{tblr}
\end{table*}


\end{document}